\documentclass[twocolumn,prd,superscriptaddress,showpacs,floatfix,%
preprintnumbers,nofootinbib]{revtex4}

\usepackage{epsfig}

\begin{document}

\title{Self-induced conversion in dense neutrino gases: Pendulum in
flavour space}

\author{Steen Hannestad}
\affiliation{Department of Physics and Astronomy, University
of Aarhus, Ny Munkegade, DK-8000 Aarhus C, Denmark}
\affiliation{Max-Planck-Institut f\"ur Physik
(Werner-Heisenberg-Institut), F\"ohringer Ring 6, 80805 M\"unchen,
Germany}

\author{Georg G.~Raffelt}
\affiliation{Max-Planck-Institut f\"ur Physik
(Werner-Heisenberg-Institut), F\"ohringer Ring 6, 80805 M\"unchen,
Germany}

\author{G{\"u}nter Sigl}
\affiliation{APC~\footnote{UMR 7164 (CNRS, Universit\'e Paris 7,
CEA, Observatoire de Paris)} (AstroParticules et Cosmologie),
  11, place Marcelin Berthelot, F-75005 Paris, France}
\affiliation{Institut d'Astrophysique de Paris, 98bis Boulevard Arago,
 75014 Paris, France}

\author{Yvonne Y.~Y.~Wong}
\affiliation{Max-Planck-Institut f\"ur Physik
(Werner-Heisenberg-Institut),
F\"ohringer Ring 6, 80805 M\"unchen, Germany}

\date{30 August 2006, Revised version of 17 October 2006}

\preprint{MPP-2006-102}

\begin{abstract}
Neutrino-neutrino interactions can lead to collective flavour
conversion effects in supernovae and in the early universe.  We
demonstrate that the case of ``bipolar'' oscillations, where a dense
gas of neutrinos and antineutrinos in equal numbers completely
converts from one flavour to another even if the mixing angle is
small, is equivalent to a pendulum in flavour space.  Bipolar flavour
conversion corresponds to the swinging of the pendulum, which begins
in an unstable upright position (the initial flavour), and passes
through momentarily the vertically downward position (the other
flavour) in the course of its motion.  The time scale to complete one
cycle of oscillation depends logarithmically on the vacuum mixing
angle.  Likewise, the presence of an ordinary medium can be shown
analytically to contribute to a logarithmic increase in the bipolar
conversion period.  We further find that a more complex (and
realistic) system of unequal numbers of neutrinos and antineutrinos is
analogous to a spinning top subject to a torque.  This analogy easily
explains how such a system can oscillate in both the bipolar and the
synchronised mode, depending on the neutrino density and the size of
the neutrino-antineutrino asymmetry. Our simple model applies strictly
only to isotropic neutrino gasses. In more general cases, and
especially for neutrinos streaming from a supernova core, different
modes couple to each other with unequal strength, an effect that can
lead to kinematical decoherence in flavour space rather than
collective oscillations. The exact circumstances under which
collective oscillations occur in non-isotropic media remain to be
understood.
\end{abstract}


\maketitle

\section{Introduction}                        \label{sec:introduction}

The ``refractive index'' caused by a medium has a strong impact on
neutrino flavour oscillations~\cite{Wolfenstein:1977ue,
Mikheev:1986gs, Mikheev:1986wj, Fuller1987, Notzold:1987ik,
Kuo:1989qe}. This matter effect is a standard ingredient for neutrino
oscillations in laboratory experiments and in astrophysical
environments.  However, when neutrinos themselves form a significant
``background medium'' as in the early universe or in core-collapse
supernovae, the oscillation equations become nonlinear, sometimes
resulting in surprising collective phenomena. Based on Pantaleone's
key observation that neutrinos as a background medium produce a
flavour off-diagonal refractive index~\cite{Pantaleone:1992eq}, the
behaviour of dense neutrino gases was investigated in a series of
papers by Samuel, Kosteleck\'y and Pantaleone~\cite{Samuel:1993uw,
Kostelecky:1994ys, Kostelecky:1993yt, Kostelecky:1993dm,
Kostelecky:1995dt, Kostelecky:1995xc, Samuel:1996ri,
Kostelecky:1996bs, Pantaleone:1998xi}. Two classes of collective
effects were identified in these papers: we shall call them
``synchronised'' and ``bipolar'' oscillations, respectively.

Synchronised oscillations occur when the neutrino-neutrino interaction
potential is large compared to the ordinary oscillation frequencies in
vacuum or in a medium, and a sufficiently large asymmetry exists
between the neutrino and antineutrino distributions.  As a result, all
neutrinos and antineutrinos oscillate with the same frequency that is
a certain average of the ordinary oscillation frequencies. In the
spin-precession analogy of flavour oscillations, the flavour
polarisation vectors of all neutrino modes form one big spin that
precesses in a weak ``external magnetic field.'' This big spin is held
together by the strong ``internal magnetic field'' formed in flavour
space by the strong neutrino-neutrino
interaction~\cite{Pastor:2001iu}. Various qualitative and quantitative
aspects of synchronised oscillations and applications to neutrino
flavour oscillations in the early universe~\cite{Lunardini:2000fy,
Dolgov:2002ab, Wong:2002fa, Abazajian:2002qx} were studied a few years
ago, motivated by the question if neutrinos with chemical potentials
reach flavour equilibrium in the early universe before the epoch of
big-bang nucleosynthesis.

Much less attention has been paid to bipolar flavour conversions, but
very recently it has been recognised that this peculiar phenomenon
likely plays a crucial role for supernova neutrino
oscillations~\cite{Duan:2005cp, Duan:2006an, Duan:2006jv}. In the
simplest case, bipolar oscillations occur in a dense gas of equal
numbers of neutrinos and antineutrinos of the same flavour. For a
suitable mass hierarchy, even a small mixing angle can cause a
complete conversion of both neutrinos and antineutrinos to the other
flavour. We stress that a non-vanishing vacuum mixing angle is
pivotal; bipolar oscillations, although a quasi self-induced effect,
will not occur if the vacuum mixing angle strictly vanishes.

The relevant conditions for bipolar conversions probably occur for
neutrinos streaming off a collapsed supernova core where one expects a
hierarchy of average energies $\langle E_{\nu_e}\rangle<\langle
E_{\bar\nu_e}\rangle <\langle E_{\nu_x}\rangle$ with
$\nu_x=\nu_\mu,\bar\nu_\mu,\nu_\tau,\bar\nu_\tau$~\cite{Keil:2002in}.
Assuming equal luminosities for all species, the number flux of, say,
$\nu_\mu$ and $\bar\nu_\mu$ is each smaller than that of $\nu_e$ and
$\bar\nu_e$ respectively. For the inverted mass hierarchy one would
then expect bipolar oscillations driven by the ``atmospheric''
neutrino mass difference and the small mixing angle
$\Theta_{13}$.\footnote{We concentrate here on 13-mixing because
$\nu_\mu$ and $\nu_\tau$ streaming off a supernova core have equal
spectra, behave equally in ordinary matter, and are known to be nearly
maximally mixed so that the 23-mixing is not relevant in this
context. The ``solar'' mass difference is much smaller and the
corresponding hierarchy is normal so that {12}-mixing is expected to
lead to less prominent effects at larger distances from the supernova
core. In general, however, one would need to perform a three-flavour
treatment.}

Bipolar flavour conversion occurs when the neutrino-neutrino
interaction energy, $\mu=\sqrt{2}G_{\rm F}n_\nu$, exceeds a typical
vacuum oscillation frequency $\omega=\Delta m^2/2E$.  Furthermore, the
neutrino-antineutrino asymmetry must not be too large in a sense to be
quantified later.  Once these conditions are met, bipolar oscillations
take place for astonishingly small values of the mixing angle and are
nearly unaffected by the presence of an ordinary background medium,
even if it is much denser than the neutrino gas~\cite{Duan:2005cp}.
These oscillations are also unrelated to a
Mikheyev-Smirnov-Wolfenstein (MSW) resonance of the ordinary
background medium.\footnote{Of course, it has long been recognised
that the high density of neutrinos near a supernova neutrino sphere
could lead to refractive effects comparable to those of the ordinary
medium, and nonlinear effects could play an important
role~\cite{Pantaleone:1994ns, Qian:1994wh, Sigl:1994hc}.  The
conclusion of these early works appears to be that the
neutrino-neutrino term causes a small shift of the oscillation
parameters where an MSW resonance occurs and hence a small correction
to the ordinary matter effect. Subsequent numerical studies of
nonlinear effects in the supernova hot bubble region actually revealed
flavour conversion for surprisingly small neutrino mass differences,
but a connection to the bipolar oscillation mode was not
made~\cite{Pastor:2002we, Balantekin:2004ug}.}

The notion of ``bipolar oscillations'' originates from the claimed
numerical observation that the flavour polarisation vectors of all
neutrinos and of all antineutrinos align together to form two ``block
spins'' that evolve separately.  Indeed, the analytic descriptions of
Refs.~\cite{Kostelecky:1995dt, Samuel:1996ri} were based on the study
of a system of one neutrino and one antineutrino polarisation vector
that were taken to represent, respectively, the complete ensemble of
neutrinos and antineutrinos.  We find that this description of bipolar
oscillations is incorrect and actually has never been explicitly
demonstrated in the literature.  In fact, each mode of the neutrino
and antineutrino ensemble evolves differently and the term ``bipolar''
is a misnomer. On the other hand, the behaviour {\it is\/} bipolar in
the sense that neutrinos and antineutrinos oscillate in ``opposite
directions'' and thus form two separate cohorts, even if they do not
form two block spins. Therefore, we use the term ``bipolar'' to
describe this collective phenomenon.

The newly recognised role of bipolar oscillations represents a
change of paradigm for supernova neutrino oscillations.  Inspired by
these exciting developments we turn to an analytic study of bipolar
oscillations with the aim of providing a simple qualitative and
quantitative understanding of the salient features of this puzzling
effect. Analytic solutions for certain cases have already been
provided in the literature~\cite{Kostelecky:1995dt, Samuel:1996ri}.
However, a judicious choice of variables enables us to write the
equations of motion in the form of an ordinary pendulum. This
picture allows one to grasp the salient features of the bipolar
phenomenon at a single glance.  Moreover, it allows one to calculate
explicitly the dependence of the bipolar oscillation period on the
vacuum mixing angle and on the density of an ordinary background
medium.

When the fluxes of neutrinos and antineutrinos are not equal (as
expected for supernova neutrinos), the equations for the flavour
pendulum remain the same, but the asymmetric initial conditions
imply the presence of an inner angular momentum (i.e., spin) of the
pendulum: the system is equivalent to a spinning top subject to a
torque. If the top is not spinning, we simply recover the motion of
an ordinary spherical pendulum (i.e., bipolar behaviour of a
symmetric $\nu$-$\bar\nu$ system).  Otherwise the motion is more
complicated. If the spin is sufficiently large, the top precesses in
the force field exerting the torque (i.e., synchronised
oscillations). If the spin is too small, the top wobbles or even
completely turns over (i.e., bipolar behaviour of an asymmetric
system).

In the most general case, however, different neutrino modes have
different energies and, in a non-isotropic medium such as that
encountered by neutrinos streaming off a supernova core, couple to
each other with different strengths (``multi-angle case''). In this
situation kinematical decoherence rather than collective oscillations
can obtain and is an unavoidable outcome certainly in the simplest
system of equal numbers of neutrinos and antineutrinos. However, in
the more realistic case of unequal neutrino and antineutrino fluxes in
conjunction with a slowly varying effective neutrino density,
collective oscillations often still obtain. The exact criteria that
determine if kinematical decoherence or collective oscillations occur
remain to be understood.

We begin in Sec.~\ref{sec:oscillator} with the simplest bipolar system
consisting of one polarisation vector for the neutrinos and one for
the antineutrinos and establish the equivalence to a flavour
pendulum. We then study analytically the impact of an ordinary
background medium in Sec.~\ref{sec:matter}. In Sec.~\ref{sec:density}
we allow the neutrino density to vary and show that this effect is
crucial for the nearly complete flavour conversion in supernovae. In
Sec.~\ref{sec:asymmetric} we consider a system with an initial
$\nu$-$\bar\nu$ asymmetry and show its equivalence to a spinning top
subject to a torque. We apply our insights to explain the salient
features of flavour conversion of the neutrinos streaming off a
supernova core. In Sec.~\ref{sec:manymodes} we consider a system of
many modes and discuss the conditions under which it is equivalent to
the simple bipolar system.  In Sec.~\ref{sec:quantum} we discuss the
possibility of flavour conversion initiated by quantum fluctuations,
in the absence of flavour mixing.  A summary of our findings is given
in Sec.~\ref{sec:discussion}.

\section{Basic Bipolar System}                  \label{sec:oscillator}

\subsection{Equations of motion}
\label{subsec:eom}

We begin with the simplest bipolar system initially composed of equal
densities of pure $\nu_e$ and $\bar\nu_e$. All of them are taken to
have equal energies so that the vacuum oscillation frequencies are the
same for all modes. We describe the flavour content of these ensembles
with polarisation vectors in flavour space ${\bf P}$ and $\bar{\bf
P}$, where overbarred quantities refer to antiparticles here and
henceforth.  Without loss of generality we take these vectors to have
unit length.  As usual, the $z$-component of the polarisation vector
represents the flavour content of the ensemble, i.e., the survival
probability of $\nu_e$ at time $t$ is $\frac{1}{2}[1+P_z(t)]$. We take
the positive $z$ direction to represent the electron flavour so that
both ${\bf P}$ and $\bar{\bf P}$ are initially unit vectors in the
$z$-direction.

In the absence of ordinary matter the general equations of motion
Eq.~(\ref{eq:matter10}) are
\begin{eqnarray}\label{eq:eom3}
 \partial_t{\bf P}&=&\left[+\omega{\bf B}
 +\mu\left({\bf P}-\bar{\bf P}\right)\right]\times{\bf P}\,,
 \nonumber\\
 \partial_t\bar{\bf P}&=&\left[-\omega{\bf B}
 +\mu\left({\bf P}-\bar{\bf P}\right)\right]\times\bar{\bf P}\,,
\end{eqnarray}
where $\omega>0$ is the vacuum oscillation frequency, $\mu=\sqrt{2}G_F
n_\nu$ represents the strength of the $\nu$-$\nu$ interaction, i.e.,
the density of the neutrino gas, and ${\bf
B}=(\sin2\theta_0,0,-\cos2\theta_0)$ with vacuum mixing angle
$\theta_0$.

A mixing angle close to zero corresponds to the normal mass hierarchy
in which $\nu_e$ is essentially identical with the lower mass state,
while $\theta_0$ near $\pi/2$ corresponds to the inverted hierarchy
with $\nu_e$ residing largely in the heavier mass state.  In the
latter case we will also use the notation
\begin{equation}
\tilde\theta_0=\pi/2-\theta_0\,.
\end{equation}
Therefore, whenever $\tilde\theta_0$ appears as a small quantity it
signifies directly that we are in an inverted mass situation.
Alternatively, one
could restrict the vacuum mixing angle to $0\leq\theta_0\leq\pi/4$ and
switch to the inverted hierarchy with the replacement
$\omega\to-\omega$. However, in our treatment it is more natural to
keep $\omega$ always positive and extend the range of mixing angles to
$0\leq\theta_0\leq\pi/2$.

\begin{figure}[t]
\begin{center}
\epsfig{file=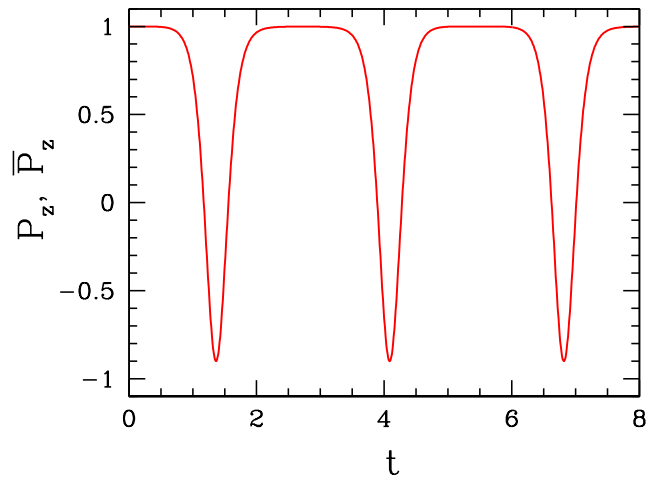,width=8.5cm}
\end{center}
\caption{Evolution of $P_z$ and $\bar{P}_z$ for the system of
  equations Eq.~(\ref{eq:eom3}) with $\tilde\theta_0=0.01$ (i.e.,
  inverted hierarchy), $\omega=1$, and strong neutrino-neutrino
  interaction $\mu=10$. (This figure is essentially identical to
  Fig.~3 of Ref.~\cite{Pastor:2001iu}.)}\label{fig:firstexample}
\end{figure}

In order to illustrate the phenomena we wish to study we show in
Fig.~\ref{fig:firstexample} the evolution of $P_z$ for a vacuum mixing
angle near $\pi/2$, corresponding to an inverted mass hierarchy. At
first ${\bf P}$ hardly moves at all, but after some time it flips
almost completely. Therefore, even a very small mixing angle leads to
complete flavour conversion. Of course, this simple system is periodic
so that the motion then reverses itself. The behaviour of $\bar{P}_z$
is identical to that of $P_z$.  In other words, both $\nu_e$ and
$\bar\nu_e$ convert simultaneously to $\nu_\mu$ and $\bar\nu_\mu$.

This evolution can be understood from Eq.~(\ref{eq:eom3}). Initially
the difference of the polarisation vectors
\begin{equation}\label{eq:Ddef}
{\bf D}={\bf P}-\bar{\bf P}
\end{equation}
vanishes so that there is no neutrino-neutrino effect. When the
polarisation vectors ${\bf P}$ and $\bar{\bf P}$ begin to precess in
opposite directions around ${\bf B}$, a non-zero ${\bf D}$ develops in
the $y$-direction orthogonal to ${\bf B}$.  Both ${\bf P}$ and
$\bar{\bf P}$ then tilt around ${\bf D}$, leading to a complete
flavour reversal (inverted hierarchy) or to small oscillations (normal
hierarchy).

Observe that ${\bf P}$ and $\bar{\bf P}$ behave symmetrically and
their $z$-components develop identically.  This suggests that
instead of these polarisation vectors one should use their sum and
difference vectors as independent variables, i.e., ${\bf D}$ as
defined in Eq.~(\ref{eq:Ddef}), and\footnote{Our vector
${\bf S}$ corresponds to ${\bf S_-}$ of Ref.~\cite{Duan:2005cp},
whereas ${\bf D}$ corresponds to their ${\bf S_+}$, i.e., what
Ref.~\cite{Duan:2005cp} calls a sum is what we call a difference,
and vice versa.  This reversal of roles arises because we work with
polarisation vectors, while Ref.~\cite{Duan:2005cp}  uses ``neutrino
flavour isospin (NFIS)'' vectors. We discuss the advantages and
disadvantages of these languages in Sec.~\ref{sec:quantum}.}
\begin{equation}\label{eq:Sdef}
{\bf S}={\bf P}+\bar{\bf P}\,.
\end{equation}
The $z$-component of ${\bf S}$ quantifies the flavour content of the
combined $\nu_e$ and $\bar\nu_e$ ensemble. The equations of motion for
the ${\bf S}$ and ${\bf D}$ vectors are
\begin{eqnarray}\label{eq:eom4}
 \dot{\bf S}&=&\omega{\bf B}\times{\bf D}
 +\mu{\bf D}\times{\bf S}\,,
 \nonumber\\
 \dot{\bf D}&=&\omega{\bf B}\times{\bf S}\,.
\end{eqnarray}
The first line of Eq.~(\ref{eq:eom4}) suggests yet another vector to
describe the ensemble,
\begin{equation}\label{eq:q}
{\bf Q}={\bf S}-\frac{\omega}{\mu}{\bf B}\,.
\end{equation}
For strong neutrino-neutrino interactions ($\mu/\omega\gg 1$) we can
think of ${\bf Q}$ as identical to ${\bf S}$.

Since $\dot {\bf B}=0$, it follows that $\dot{\bf S}=\dot{\bf Q}$.
With ${\bf B}\times{\bf Q}={\bf B}\times{\bf
S}$, the equations of motion are now
\begin{eqnarray}\label{eq:eom5}
 \dot{\bf Q}&=&\mu{\bf D}\times{\bf Q}\,,
 \nonumber\\
 \dot{\bf D}&=&\omega{\bf B}\times{\bf Q}\,.
\end{eqnarray}
Clearly, the length of ${\bf Q}$ is conserved and stays at its initial
value
\begin{equation}\label{eq:modq}
Q=|{\bf Q}|=
\left[4+\left(\frac{\omega}{\mu}\right)^2
+4\,\frac{\omega}{\mu}\cos2\theta_0\right]^{1/2}\,,
\end{equation}
where we have used $|{\bf B}|^2=1$, and the initial values
$|{\bf S}|^2=4$ and ${\bf B}\cdot{\bf S}=-2\cos2\theta_0$.

\subsection{Spherical pendulum}

The vector ${\bf Q}$ in flavour space plays the role of a spherical
pendulum in that its length is conserved so that it can move only on a
sphere of radius $Q$. In this picture, the role of the different
quantities is most easily understood if we consider the total energy
of the system,
\begin{equation}\label{eq:ham3}
H=\omega{\bf B}\cdot {\bf Q}+\frac{\mu}{2}{\bf D}^2\,,
\end{equation}
up to a constant.  The first term is the potential energy of the
pendulum in a homogeneous force field represented by $\omega {\bf B}$.
The second term is the kinetic energy, with ${\bf D}$ playing the role
of the pendulum's orbital angular momentum. Observing that ${\bf
D}\cdot{\bf Q}= -(\omega/\mu){\bf D}\cdot{\bf B}$ is constant due to
Eq.~(\ref{eq:eom5}) and thus zero in our case ${\bf D}(0)=0$, the
first line in Eq.~(\ref{eq:eom5}) implies
\begin{equation}\label{eq:angmomentum1}
{\bf D}=\frac{1}{\mu}\,\frac{{\bf Q}\times\dot{\bf Q}}{Q^2}\,,
\end{equation}
and hence ${\bf D}^2=\mu^{-2}\,\dot{\bf Q}^2/{Q^2}$.  The scale of the
potential energy is set by the vacuum neutrino oscillation frequency
$\omega$, whereas $I=\mu^{-1}$ is to be identified with the moment of
inertia.  The latter should be compared with $I=m\ell^2$ for an
ordinary mass suspended by a string of length $\ell$. The role of
inertia in the pendulum analogy is played by the {\it inverse\/}
strength of neutrino-neutrino interaction!

\subsection{Plane pendulum}

Our initial conditions ${\bf P}(0)= \bar{\bf P}(0)=(0,0,1)$ imply
${\bf D}(0)=0$. The pendulum's subsequent oscillations are confined in
a plane defined by ${\bf B}$ and the $z$-axis. Therefore, the problem
reduces to solving for the motion of the tilt angle $\varphi$ of ${\bf
Q}$ relative to the $z$-axis.  Writing ${\bf
Q}=Q\,(\sin\varphi,0,\cos\varphi)$, we find
\begin{eqnarray}\label{eq:eom6}
 \dot\varphi&=&\mu D\,,
 \nonumber\\
 \dot D&=&-\omega Q \sin(\varphi+2\theta_0)\,.
\end{eqnarray}
Equation~(\ref{eq:eom6}) can be further simplified to
\begin{equation}\label{eq:eom6a}
\ddot\varphi=-\kappa^2\,\sin(\varphi+2\theta_0)\,,
\end{equation}
where
\begin{equation}
\kappa^2=\omega\mu Q\,.
\end{equation}
The inverse of $\kappa$ is the characteristic time scale for the
bipolar evolution.  In the limit of strong neutrino-neutrino coupling,
$Q\approx 2$ and hence $\kappa \approx\sqrt{2\omega\mu}$. In the
opposite limit ($\mu\ll\omega$), we have $Q\approx \omega/\mu$ so that
$\kappa\approx\omega$. In this latter case the characteristic
frequency of the system is the vacuum oscillation frequency $\omega$,
and ${\bf P}$ and $\bar{\bf P}$ oscillate independently.

The equations of motion (\ref{eq:eom6}) follow directly from the
classical Hamiltonian for a simple pendulum
\begin{equation}\label{eq:ham1}
H(\varphi,D)=
\frac{\kappa^2}{\mu}\bigl[1-\cos(\varphi+2\theta_0)\bigr]
+\frac{1}{2}\,\mu\,D^2\,,
\end{equation}
where $\varphi$ is a coordinate and $D$ its canonically conjugate
momentum. Indeed, the equations (\ref{eq:eom6}) are but
$\dot\varphi=\partial H/\partial D$ and $\dot D=-\partial H/\partial
\varphi$.

Assuming a small vacuum mixing angle $\theta_0$ and a small excursion
angle $\varphi$ of the pendulum, the potential can be expanded
\begin{eqnarray}\label{eq:potential}
V(\varphi)&=&\kappa^2\,
\left[1-\cos(\varphi+2\theta_0)\right]\nonumber\\
&=&\frac{\kappa^2}{2}\,
\left(\varphi+2\theta_0\right)^2+\ldots\,.
\end{eqnarray}
In this case the system is equivalent to a harmonic oscillator with
frequency $\kappa$. On the other hand, for angles near $\pi$ so that
$|\varphi+2\theta_0-\pi|\ll 1$, we get the same expansion but with a
negative sign; the system corresponds to an inverted harmonic
oscillator.

\subsection{Bipolar flavour conversion}

\subsubsection{Normal hierarchy}

Consider the small mixing limit in the normal hierarchy.  Here, the
initial condition $\varphi(0)\approx -(\omega/\mu Q) \ 2 \theta_0 \geq
-2 \theta_0$ puts $\varphi$ near the minimum of the potential
$V(\varphi)$ at $t=0$.  Since $\dot\varphi(0)=0$, the system remains
trapped inside the cosine potential well, oscillating about the
minimum $\varphi_{\rm min}=-2 \theta_0$ with amplitude $(1-\omega/\mu
Q) \ 2 \theta_0$ and frequency $\kappa$.  In terms of $P_z$ and
$\bar{P}_z$, any departure from the initial $P_z=\bar{P}_z=1$ is at
most second order in $\theta_0$.  No dip features develop in this
scenario.

\subsubsection{Inverted hierarchy with arbitrary $\mu$}

Using the relation $\tilde\theta_0=\pi/2-\theta_0$, the potential
(\ref{eq:potential}) can be written as
\begin{eqnarray}\label{eq:potential2}
V(\varphi)&=&\kappa^2\,
\left[1+\cos(\varphi-2\tilde\theta_0)\right]\nonumber\\
&=&-\frac{\kappa^2}{2}\,
\left(\varphi-2\tilde\theta_0\right)^2+\ldots\,
\end{eqnarray}
in the inverted hierarchy. Depending on the strength of the
neutrino-neutrino interactions (i.e., the ratio $\mu/\omega$),
$\varphi$ can take on a range of initial values from
$\varphi(0)\approx 0$ to $2 \tilde\theta_0 - \pi$ in the small
$\tilde\theta_0$ limit.  In other words, the evolution of the system
begins with $\varphi$ sitting near the maximum of the potential
$V(\varphi)$ if $\mu/\omega \geq 1$, or, for smaller values of
$\mu/\omega$, somewhere further down the slope.

Since $\dot\varphi(0)=0$, the form of the potential $V(\varphi)$
guarantees that $\varphi$ {\it always} rolls to the minimum
$\varphi_{\rm min} \approx -\pi$ in the small $\tilde\theta_0$ limit.
Evaluating for $P_z$ and $\bar{P}_z$ at $\varphi=\varphi_{\rm min}$,
one finds
\begin{equation}
P_z|_{\varphi_{\rm min}}=\bar{P}_z|_{\varphi_{\rm min}}
\approx \left\{\begin{array}{ll}
        \omega/\mu -1, &\qquad \omega < 2 \mu, \\
    1, & \qquad \omega \geq 2 \mu.\end{array} \right.\,
\end{equation}
Thus complete flavour conversion, i.e., $P_z =\bar{P}_z = -1$, is only
possible in the strong neutrino-neutrino coupling limit $\mu/\omega
\gg 1$. A partial conversion can be achieved for comparable $\mu$ and
$\omega$.  No conversion occurs for $\omega \geq 2 \mu$, which
remains always in the small oscillations regime.

\subsubsection{Inverted hierarchy with $\mu\geq\omega$}

We now focus on $\mu/\omega \geq 1$.  The initial condition
$\varphi(0) \approx -(\omega/\mu Q) \ 2 \tilde\theta_0$ puts $\varphi$
near the maximum of the potential $V(\varphi)$ at $t=0$.  Since
$\dot\varphi(0)=0$, and $\varphi(0)$ and $2\tilde\theta_0$ are both
small, the motion of $\varphi$ begins slowly, which explains the long
plateau phases between dips in Fig.~\ref{fig:firstexample}. The dips
themselves correspond to the crossing of the anharmonic potential once
$\varphi$ grows to the order unity.  The duration of the dips is fixed
by the crossing time of the anharmonic potential and thus is of order
$\kappa^{-1}$.  The duration of the plateau, on the other hand, is
determined by the smallness of the mixing angle $\tilde\theta_0$.  If
$\tilde\theta_0$ is exactly zero, i.e., no mixing, then there is no
motion and $\varphi$ sits at the exact maximum of the potential
forever.

In order to estimate the time it takes for the polarisation vectors to
flip in the small $\tilde\theta_0$ limit, we return to the equation of
motion for this case,
\begin{equation}
\ddot\varphi=\kappa^2\sin(\varphi-2\tilde\theta_0),
\end{equation}
with $\tilde\theta_0\ll 1$. As long as $\varphi$ is small this is
equivalent to
\begin{equation}\label{eq:invertosci}
\ddot\varphi=\kappa^2(\varphi-2\tilde\theta_0)\,.
\end{equation}
Using the initial conditions $\varphi(0) = -(\omega/\mu Q) \ 2
\tilde\theta_0$ and $\dot\varphi(0) = 0$, this equation is solved by
\begin{equation}
 \varphi(t)=2\tilde\theta_0
 \left[1-\left(1+\frac{\omega}{\mu Q} \right)
 \cosh(\kappa t)\right]\,.
\end{equation}
Initially $\varphi(t)-\varphi(0)=-\tilde\theta_0(1+\omega/\mu
Q)(\kappa t)^2$, but at $t$ of order $\kappa^{-1}$ turns to
exponential growth when $\varphi$ has become of order
$\tilde\theta_0$. Therefore, the time it takes for $\varphi$ to grow
to order unity, or equivalently, the half period of the bipolar
motion is
\begin{equation}
\tau_{\rm bipolar}\approx-\kappa^{-1}
\ln[\tilde\theta_0\ (1+\omega/\mu Q)]\,.
\end{equation}
Therefore, the duration of the plateau phases in
Fig.~\ref{fig:firstexample} or the time between dips scales
logarithmically with the small vacuum mixing angle.

\subsection{Which mass hierarchy?}

We demonstrated that, assuming small mixing, complete flavour
conversion occurs for the inverted mass hierarchy, while small
oscillations occur for the normal hierarchy.  However, this applies
only if the initial ensemble consists of $\nu_e$ and $\bar\nu_e$. If
the initial ensemble had consisted instead of $\nu_\mu$ and
$\bar\nu_\mu$, then the situation would be reversed so that large
flavour conversions would occur for the normal hierarchy.  Put another
way, the unstable case is when the initial ensemble consists of that
flavour which is dominated by the heavier mass eigenstate.  This
symmetry between the neutrino flavours is an important difference to
flavour conversion caused by the MSW effect that occurs for the normal
mass hierarchy independently of the flavour of the initial state.

\subsection{Which flavour conversion?}

We have used the term ``flavour conversion'' loosely to describe the
simultaneous conversion of equal numbers of $\nu_e$ and $\bar\nu_e$ to
equal numbers of $\nu_\mu$ and $\bar\nu_\mu$. Of course, the net
flavour lepton number of the initial state vanishes and remains so at
all times. Therefore, the ``conversion'' we are considering in the
bipolar context does not violate flavour lepton number, but, rather,
we should think of it as a coherent pair process of the form
$\nu_e\bar\nu_e\to\nu_\mu\bar\nu_\mu$.

If the initial system is asymmetric with a net electron lepton number,
i.e., ${\bf P}$ and $\bar{\bf P}$ are not initially identical, this
difference is quantified by our vector ${\bf D}$, whose projection in
the $z$-direction represents the net flavour lepton number.  From
Eq.~(\ref{eq:eom5}) we observe that ${\bf D}\cdot{\bf B}$ is
conserved.  In other words, there is no net conversion of flavour
beyond what is caused by ordinary vacuum oscillations, and the net
lepton numbers of vacuum mass eigenstates are strictly conserved. This
statement is independent of the strength of $\mu$, and applies
irrespective of the asymmetric system being in the synchronised or
bipolar regime. It also applies to the multi-mode system described in
a later section.

\section{Background Matter}
\label{sec:matter}

An ordinary background medium has little impact on the bipolar flavour
conversion~\cite{Duan:2005cp, Duan:2006an, Duan:2006jv}. This
surprising observation runs against the intuition that in a medium the
mixing angle should be suppressed. We have already found that the time
scale for bipolar flavour conversion depends only logarithmically on
the vacuum mixing angle. One would perhaps expect this time scale to
depend also logarithmically on the matter density.

To investigate this case we include matter effects caused by charged
leptons in  the equations of motion,
\begin{eqnarray}\label{eq:matter1}
 \partial_t{\bf P}&=&\left[+\omega{\bf B}+\lambda{\bf L}
 +\mu\left({\bf P}-\bar{\bf P}\right)\right]\times{\bf P}\,,
 \nonumber\\
 \partial_t\bar{\bf P}&=&\left[-\omega{\bf B}+\lambda{\bf L}
 +\mu\left({\bf P}-\bar{\bf P}\right)\right]\times\bar{\bf P}\,.
\end{eqnarray}
In the absence of other terms, ${\bf P}$ and $\bar{\bf P}$ would
precess around ${\bf L}$ in the same direction with a frequency
$\lambda$. Therefore, it was noted in Refs.~\cite{Duan:2005cp,
Duan:2006an} that the equations simplify if we study them in a frame
co-rotating around the ${\bf L}$-direction, i.e., around the
$z$-direction.

In the co-rotating frame, the equations of motion for this system
take the form of our original equations~(\ref{eq:eom3}),
except that ${\bf B}$ is now time dependent,
rotating around the $z$-direction with frequency $-\lambda$,
\begin{equation}\label{eq:bt}
  {\bf B}=\left(\matrix{\sin(2\theta_0)\cos(-\lambda t)\cr
  \sin(2\theta_0)\sin(-\lambda t)\cr
  -\cos(2\theta_0)\cr}\right).
\end{equation}
If this rotation is faster than all other frequencies one would
na\"{\i}vely expect that the transverse components of ${\bf B}$
average to zero, leaving us with $\langle {\bf B}\rangle$ along the
$z$-axis, i.e., an effectively vanishing mixing angle and no flavour
conversion. However, as it turns out, there remains a net effect on
the polarisation vectors, and bipolar flavour conversions occur after
all!

To understand this case quantitatively and qualitatively we consider
the full set of equations~(\ref{eq:eom3}), keeping in mind the
time dependence of the ${\bf B}$ vector in the co-rotating frame
[cf.\ Eq.~(\ref{eq:bt})].  We define a new vector ${\bf Q}$
as per Eq.~(\ref{eq:q}).  However, since ${\bf B}$ is now
time-dependent, the new equations of motion have a slightly more
complex structure than Eq.~(\ref{eq:eom5}),
\begin{eqnarray}\label{eqsofmotion}
 \dot{\bf Q}&=& \mu{\bf D}\times{\bf Q}
 -\frac{\omega}{\mu} \dot{\bf B},
 \nonumber\\
 \dot{\bf D}&=& \omega{\bf B}\times{\bf Q}\,,
\end{eqnarray}
with
\begin{equation}
\dot{\bf B} = \lambda \sin(2 \theta_0)
\left(\matrix{\sin(-\lambda t)\cr
    -\cos(-\lambda t) \cr
    0 \cr} \right).
\end{equation}
Thus ${\bf Q}$ is not strictly conserved; its length $|{\bf Q}|$ will,
in the small $\theta_0$ limit, exhibit order $\theta_0$ fluctuations
around some mean value given by Eq.~(\ref{eq:modq}).

Note that ${\bf D}\cdot{\bf B}$ too  is no longer exactly conserved.
However, this non-conservation is likely to become relevant only in
the vacuum oscillation dominated regime and for significant vacuum
mixing angles: Since the non-constant part of ${\bf B}$ is
proportional to $\sin(2\theta_0)$ and oscillates with frequency
$\lambda$, while, according to Eq.~(\ref{eqsofmotion}), ${\bf D}$
evolves with frequency $\omega$, the time varying part of ${\bf
D}\cdot{\bf B}$ is proportional to $\sin(2\theta_0)$ and will tend
to average out in the matter dominated regime we are interested in.
Of course, if the matter term varies adiabatically, a large change
of ${\bf D}\cdot{\bf B}$ in the form of the ordinary MSW effect is
possible.

We are interested in the case of small mixing in an inverted
hierarchy, so that
\begin{equation}
  {\bf B}=\left(\matrix{2\tilde\theta_0\,\cos(-\lambda t)\cr
  2\tilde\theta_0\,\sin(-\lambda t)\cr
  1\cr}\right)\,
\end{equation}
to the lowest order in $\tilde\theta_0$.
Furthermore, since we are concerned only with instances at which
the deviation of ${\bf Q}$ from the $z$-direction is small,
we can parameterise the motion of ${\bf Q}$ by two small tilt angles
$\varphi_x$ and $\varphi_y$, so that to lowest order
\begin{equation}
  {\bf Q}= Q \left(\matrix{\varphi_x\cr
  \varphi_y\cr
  1\cr}\right)\,,
\end{equation}
and $\dot{|{\bf Q}|}$ vanishes.
As a consequence of this expansion, $D_z$ and $\dot D_z$ are of
higher order.  In other words, ${\bf D}$ to lowest order does not
develop a $z$-component; it remains a vector in the $x$-$y$-plane.
One then finds a simple set of equations of motion,
\begin{eqnarray}\label{eq:matter3}
  \ddot\varphi_x&=&\kappa^2
  \left[\varphi_x-2\tilde\theta_0
  \left(1-\frac{\lambda^2}{\mu^2 Q^2} \right)\cos(\lambda t)\right]\,,
  \nonumber\\
  \ddot\varphi_y&=&\kappa^2
  \left[\varphi_y +2\tilde\theta_0
  \left(1-\frac{\lambda^2}{\mu^2 Q^2} \right)\sin(\lambda t)\right]\,,
\end{eqnarray}
with $\kappa^2=\omega \mu Q$.  Of course, had we considered the normal
hierarchy in which $B_z \approx -1$, we would have found two driven
harmonic oscillator equations instead.

Using the initial conditions
$\varphi_x(0)=-(\omega/\mu Q) \ 2 \tilde\theta_0$,
$\dot\varphi_x(0)=0$,
$\varphi_y(0)=0$, and
$\dot\varphi_y(0)=(\lambda \omega/\mu Q) \ 2 \tilde\theta_0$,
Eq.~(\ref{eq:matter3}) is solved~by
\begin{widetext}
\begin{eqnarray}
 \varphi_x(t) &=&  -2\tilde\theta_0\frac{\kappa^2}{\kappa^2+\lambda^2}
 \left[\left(1+\frac{\omega}{\mu Q}\right) \cosh(\kappa t)-
 \left(1-\frac{\lambda^2}{\mu^2 Q} \right)
 \cos(\lambda t)\right]\,,
 \nonumber\\
 \varphi_y(t)&=&-2\tilde\theta_0\frac{\kappa^2}{\kappa^2+\lambda^2}
 \left[\frac{\lambda}{\kappa}
  \left(1+\frac{\omega}{\mu Q}\right) \sinh(\kappa t)+
  \left(1-\frac{\lambda^2}{\mu^2 Q} \right)  \sin(\lambda t)\right]\,.
\end{eqnarray}
\end{widetext}
Therefore, the tilt angles have a small oscillatory motion driven by
the rotating ${\bf B}$ vector. For $t\agt\kappa^{-1}$ these
oscillatory terms no longer matter much relative to the exponentially
growing terms which scale asymptotically as $e^{\kappa t}$. The most
remarkable feature, however, is that the exponential term for
$\varphi_y$ involves an additional factor $\lambda/\kappa$ that is
absent for $\varphi_x$.

\begin{figure}[b]
\begin{center}
\epsfig{file=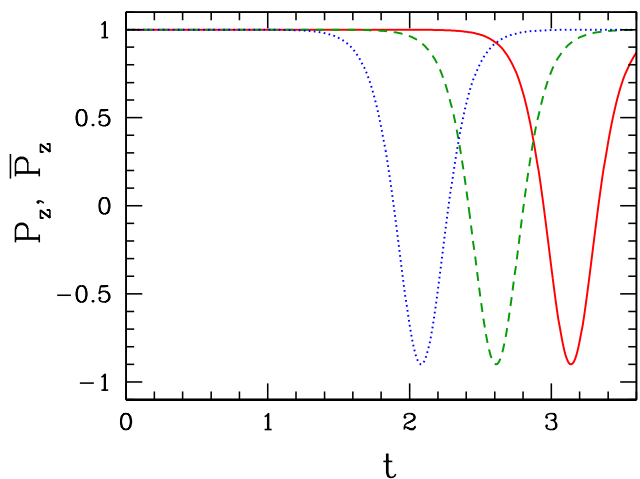,width=8.5cm}
\end{center}
\caption{Evolution of $P_z$ and $\bar{P}_z$ in several systems with
background matter described by Eq.~(\ref{eq:matter1}). The
parameters $\tilde\theta_0=0.01$ (i.e., inverted hierarchy),
$\omega=1$, and $\mu=10$ are common for all three systems.  The
blue/dotted line has $\lambda=10^2$, the green/dashed line
$\lambda=10^3$, and the red/solid line
$\lambda=10^4$.}\label{fig:pmatter}
\end{figure}

If the matter effect is small, $\lambda\ll\kappa=(\omega\mu
Q)^{1/2}$, the tilt is mostly in the $x$-direction in the
co-rotating frame.  The tilt in the $y$-direction is relatively
suppressed by a factor $\lambda/\kappa$.  In the limit $\lambda \to
0$, the co-rotating frame coincides with the ``laboratory'' frame,
and the tilt occurs exclusively in the $x$-direction as already seen
in Sec.~\ref{sec:oscillator}.  On the other hand, when the matter
effect is strong, $\lambda\gg\kappa$, the opposite applies. The tilt
is mostly in the $y$-direction in the co-rotating frame, the motion
in the $x$-direction being relatively suppressed by
$\kappa/\lambda$. We have observed this counter-intuitive behaviour
also in numerical examples where indeed ${\bf Q}$ tilts in the
$y$-direction when the matter effect is large.

To track the flavour evolution, only the $z$-component of ${\bf
Q}$ is of interest; its tilting direction is irrelevant. For
$t\gg\kappa^{-1}$, we ignore the oscillatory terms and use the
asymptotic behaviour $\cosh(\kappa t)\approx\sinh(\kappa t)\approx
\frac{1}{2} e^{\kappa t}$ so that
\begin{equation}
 \varphi(t)=\left(\varphi_x^2+\varphi_y^2\right)^{1/2}
 =\tilde\theta_0\,
 \frac{\kappa}{\left(\kappa^2+\lambda^2\right)^{1/2}}
 \,\left(1+\frac{\omega}{\mu Q}\right) \,e^{\kappa t}\,.
\end{equation}
Therefore, the time scale for flavour conversion is
\begin{equation}
  \tau_{\rm bipolar}\approx-\kappa^{-1}\ln\left[
  \tilde\theta_0\,
  \frac{\kappa}{\left(\kappa^2+\lambda^2\right)^{1/2}}
\left(1+\frac{\omega}{\mu Q} \right)
\right]\,.
\end{equation}
The presence of matter has little impact on the overall behaviour of
the bipolar system except for a logarithmic extension of $\tau_{\rm
bipolar}$.  This effect is illustrated in Fig.~\ref{fig:pmatter}.

\section{Varying Neutrino Density}
\label{sec:density}

In numerical simulations of the flavour evolution of supernova
neutrinos in the single-angle approximation, one observes almost
complete flavour conversion, apparently caused by the bipolar
effect. We have seen that a bipolar system does lead to almost
complete conversion, but also that the evolution is periodic.
Therefore, being in the bipolar regime alone does not explain complete
conversion. We have also seen that the impact of ordinary matter on
the bipolar system is negligible.  Therefore, it appears that the
decline of the neutrino density, and therefore of $\mu$, along the
neutrino flux is the likely cause of almost complete flavor
conversion.

\begin{figure}[b]
\begin{center}
\epsfig{file=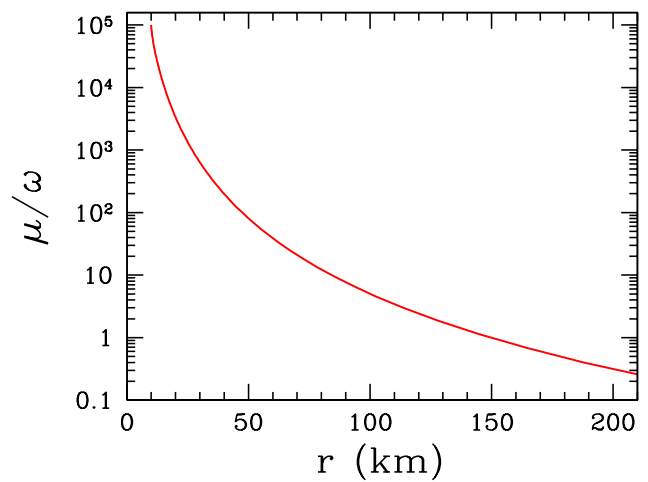,width=8.5cm}
\end{center}
\caption{Neutrino-neutrino interaction strength $\mu$ in units of
$\omega$ for our toy model of supernova neutrino
oscillations.}\label{fig:muradius}
\end{figure}

To study the impact of a time-varying (or, in a supernova,
space-varying) neutrino density in a concrete example, we assume that
all neutrinos have the same energy and oscillate in vacuum with the
frequency $\omega=0.3~{\rm km}^{-1}$, corresponding to the
``atmospheric'' neutrino mass difference for typical supernova
neutrino energies [see Eq.~(\ref{eq:wsn})]. We express all
``frequencies'' in units of km$^{-1}$ and all length scales in km as
appropriate for the supernova environment. For the $\nu$-$\nu$
interaction energy we use $\mu=0.3\times10^5~{\rm km}^{-1}$ at the
neutrino sphere ($r=10~{\rm km}$), and a dependence on the radius
given by Eq.~(\ref{eq:musn}), i.e., essentially an $r^{-4}$ scaling.
This scaling reflects both the ordinary flux dilution with $r^{-2}$,
and the degree of collinearity in the neutrino motion which introduces
approximately another factor $r^{-2}$. The quantity $\mu/\omega$ as a
function of radius is shown in Fig.~\ref{fig:muradius}.

\begin{figure}[t]
\bigskip
\begin{center}
\epsfig{file=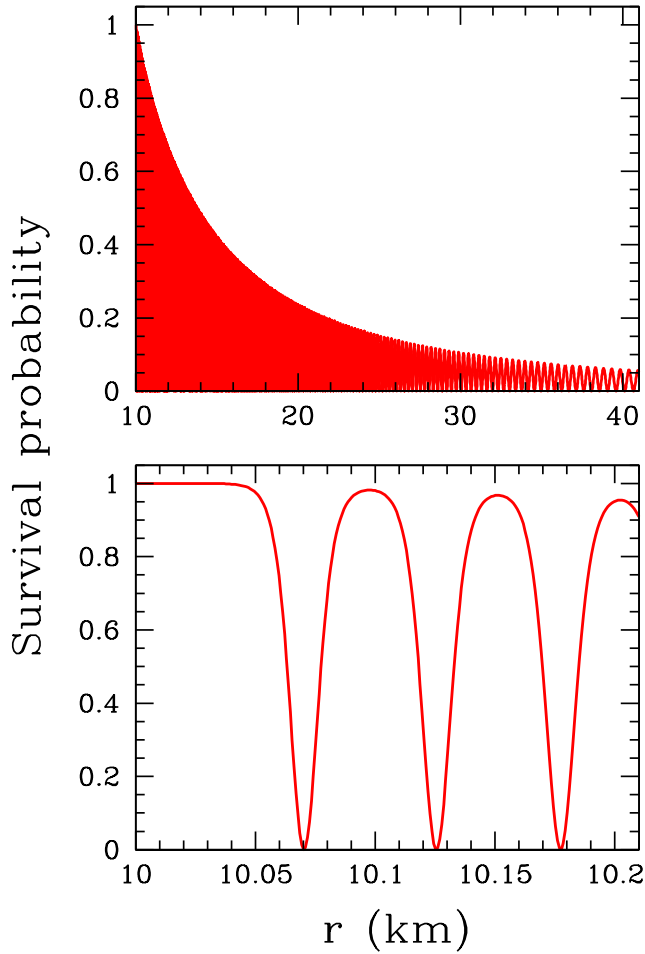,width=8.5cm}
\end{center}
\caption{Survival probabilities for $\nu_e$ or $\bar\nu_e$ in our
toy supernova model with symmetric initial
conditions.}\label{fig:SNbimodal}
\end{figure}

Furthermore, we assume that (i) we have equal fluxes of neutrinos and
antineutrinos, (ii) all of them are initially in the same flavour,
(iii) the mass hierarchy is inverted, and (iv) the mixing angle is
$\sin2\tilde\theta_0=0.001$, representing a possibly small
$\Theta_{13}$ mixing angle. We then find numerically the survival
probability of the initial flavour as shown in
Fig.~\ref{fig:SNbimodal}.

As expected, we observe in Fig.~\ref{fig:SNbimodal}
bipolar oscillations above the neutrino sphere
at $r=10$~km. Moreover, we observe that the oscillation amplitude
declines as a function of radius so that after a few tens of km we
obtain complete flavour conversion. The question then is: can the
decline of the upper envelope of the survival probability be
explained by way of the flavour pendulum?

In the pendulum language, we start at small radii with the usual full
oscillations. However, while the pendulum oscillates, we slowly reduce
$\mu$, i.e., we increase adiabatically the moment of inertia
$I=\mu^{-1}$.  Since the kinetic energy is $D^2/2I$ and the angular
momentum $D$ is conserved, an increase in $I$ corresponds to a
decrease in the kinetic energy. This is akin to a dancer who can speed
up or slow down a pirouette by changing the moment of inertia.  The
continuous decrease in $\mu$ and hence in the kinetic energy, however,
means that during subsequent swings, the pendulum will be unable to
reach its previous height.  This explains the general feature of a
declining upper envelope in the survival probability caused by a
decreasing $\mu$ (Fig.~\ref{fig:SNbimodal}).

It is important to note that the potential energy is independent
of $\mu$; it depends only on $\omega$. Therefore, our pendulum is not
like a gravitational one where the potential energy would be affected
by a changing mass.  Instead, we can imagine the bob of the pendulum
being charged and feeling the force of a homogeneous electric~field.

Let us now quantify the decline of the upper envelope in
Fig.~\ref{fig:SNbimodal}. The kinetic energy at a given time is
$T=\mu(t) D(t)^2/2$. The conservation of angular momentum, except for
the natural pendulum motion, implies that a sudden change $\Delta\mu$
at some time $t$ causes a change in kinetic energy of $\Delta
T=\Delta\mu D(t)^2/2$. For example, if we change $\mu$ by $\Delta\mu$
when the pendulum swings past its lowest point at which the kinetic
energy is maximal, then the relative change is $\Delta T_{\rm
max}/T_{\rm max}=\Delta\mu/\mu$. However, $\mu$ decreases slowly
compared to the oscillation period so that we may assume a linear
decline. Therefore, the change of $T_{\rm max}$ occurs over the entire
oscillation period and thus must be weighted with a factor
proportional to $D(t)^2$ over one oscillation period. If the
oscillation is approximately harmonic, we have
$D(t)^2\propto\sin^2(\kappa t)$ with $\kappa$ the pendulum's natural
frequency. The average of $\sin^2(\kappa t)$ is $1/2$. Therefore, if
over one period $\mu$ decreases by a factor $(1-\epsilon)$ where
$\epsilon\ll 1$, then $\Delta T_{\rm max}/T_{\rm max}=-\epsilon/2$.
In other words, $T_{\rm max}$ is reduced by a factor
$(1-\epsilon/2)=(1-\epsilon)^{1/2}$ so that $T_{\rm max}\propto
\mu^{1/2}$.

\begin{figure}[b]
\begin{center}
\epsfig{file=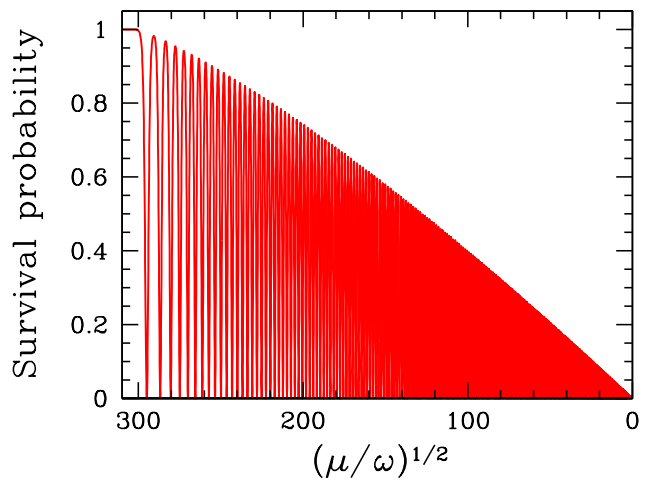,width=8.5cm}
\end{center}
\caption{Survival probability for $\nu_e$ or $\bar\nu_e$ as in
Fig.~\ref{fig:SNbimodal}, here plotted as a function of
$(\mu/\omega)^{1/2}$.}\label{fig:mudecline}
\end{figure}

The maximal kinetic energy equals the maximal potential energy minus
its minimum, achieved within one oscillation. The potential energy
normalised such that its minimum is zero, $\omega\left(2+{\bf
B}\cdot{\bf Q}\right)$ for small mixing angles and for large
$\mu/\omega$, gives us the projection of the summed polarisation
vector ${\bf S}$ on the flavour axis. Therefore, the upper envelope of
the survival probability in Fig.~\ref{fig:SNbimodal} should scale with
$\mu^{1/2}$. In Fig.~\ref{fig:mudecline} we show the same survival
probabilities with $(\mu/\omega)^{1/2}$ as a radial coordinate.  The
decline is indeed nearly linear, especially for small-amplitude
oscillations (toward the right side of the plot) where the pendulum
oscillations are more nearly~harmonic.

Our pendulum analogy elegantly explains the most puzzling feature of
the bipolar supernova neutrino oscillations, i.e., that they actually
lead to flavour conversion rather than permanent bipolar
oscillations. We are simply seeing the relaxation of the pendulum to
its downward rest position as kinetic energy is extracted by the
reduction of the neutrino-neutrino interaction potential and thus the
increase of the pendulum's inertia.

\section{Neutrino Asymmetry}
\label{sec:asymmetric}

\subsection{Realistic supernova example}
\label{subsec:realistic}

\begin{figure}[t]
\begin{center}
\epsfig{file=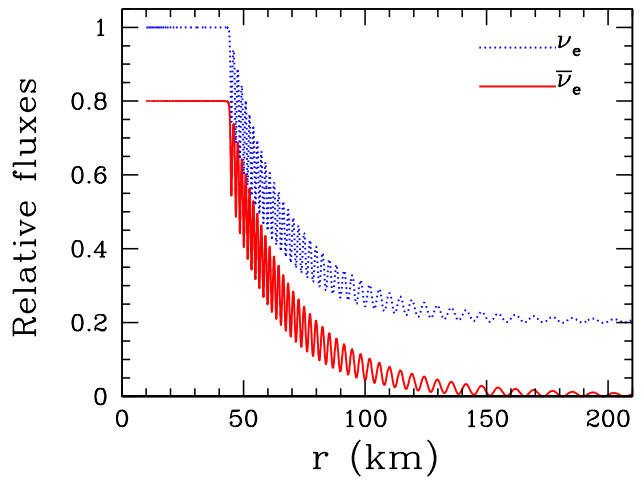,width=8.5cm}
\end{center}
\caption{Relative fluxes of $\nu_e$ (blue/dotted) and $\bar\nu_e$
(red/solid) in our toy supernova model with 20\% fewer antineutrinos
than neutrinos and $\sin2\tilde\theta_0=0.001$.}\label{fig:SN2}
\end{figure}

The behaviour of neutrinos streaming off a supernova core looks,
however, rather different from this simple picture. Besides the
dependence of $\mu$ on the radius, another crucial feature is the
initial neutrino-antineutrino asymmetry; the number flux of $\nu_e$ is
the largest, that of $\bar\nu_e$ smaller, and those of the other
species yet smaller but equal to one another. We represent this
situation by the initial conditions $P_z(0)=1$ and $\bar P_z(0)=0.8$.
The equations of motion for this system are simply those given in
Sec.~\ref{subsec:eom}.  Solving them numerically yields the relative
$\nu_e$ and $\bar\nu_e$ fluxes shown in Fig.~\ref{fig:SN2}.  Plotting
relative fluxes instead of survival probabilities makes conservation
of the net flavour-lepton flux evident.

Observe that the initial flavour-lepton asymmetry is conserved so that
there remains a net flux of $\nu_e$ originally set at
$r=10$~km. Otherwise there is complete flavour conversion over a
length scale given by the decrease of $\mu$ as a function of
radius. Similar results are found in detailed numerical studies within
the ``one-angle approximation'' where the neutrino interaction
strength is taken equal for all modes (Fig.~8c of
Ref.~\cite{Duan:2006an} and private communication by S.~Pastor and
R.~Tom\`as based on the numerical scheme of
Ref.~\cite{Pastor:2002we}). Changing the vacuum mixing angle and
adding normal matter causes only the minor logarithmic changes
predicted earlier.

The behaviour of the neutrino ensemble between the neutrino sphere
at 10~km and $r\approx45$~km is explained by synchronised
oscillations due to the neutrino-antineutrino asymmetry and the
large value of $\mu$ in this region. Beyond this we enter the
bipolar regime out to about 200~km when vacuum oscillations take
over. In other words, the flavour evolution of neutrinos
streaming off a supernova core are determined by a transition from
synchronised oscillations at small radii (large $\mu$), bipolar
oscillations at intermediate radii (intermediate $\mu$), and
ordinary vacuum oscillations at large radii (small $\mu$) as first
stressed in Ref.~\cite{Duan:2005cp}. If ordinary matter is included,
it affects the synchronised region at small radii in the usual way
by making the effective mixing angle smaller, and likewise at large
radii where no collective effects occur. It is only in
the intermediate, bipolar oscillation regime
where ordinary matter has no significant impact on the system.
Confusingly, the bipolar
behaviour does not correspond to the limit of large or
small neutrino densities; it corresponds to {\it intermediate\/}
densities~\cite{Duan:2005cp}.

\begin{figure}[t]
\begin{center}
\epsfig{file=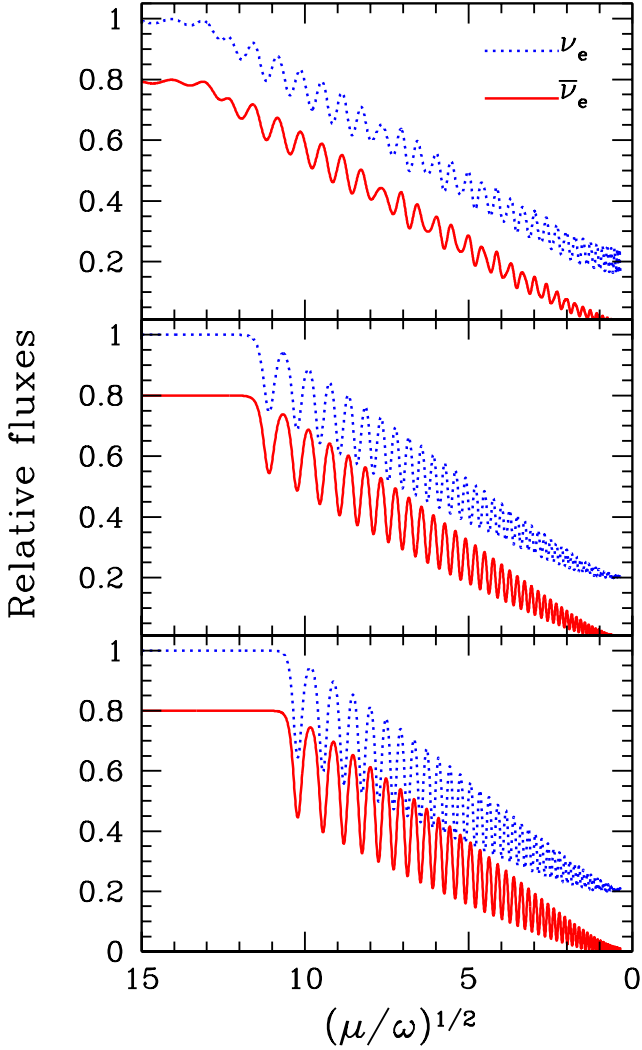,width=8.5cm}
\end{center}
\caption{Relative fluxes of $\nu_e$ (blue/dotted) and $\bar\nu_e$
(red/solid) as in Fig.~\ref{fig:SN2}, here using $(\mu/\omega)^{1/2}$
as a radial coordinate. {\it Top:\/} Vacuum mixing angle
$\sin2\tilde\theta_0=0.1$. {\it Middle:\/}
$\sin2\tilde\theta_0=10^{-3}$ as in Fig.~\ref{fig:SN2}.  {\it
Bottom:\/} $\sin2\tilde\theta_0=10^{-5}$.}\label{fig:musurvival}
\end{figure}

We have already explained the decline of the upper envelope of these
curves in the bipolar regime which should scale as $\mu^{1/2}$. To
confirm this behaviour once more we show in
Fig.~\ref{fig:musurvival} (middle panel) the relative fluxes
of Fig.~\ref{fig:SN2} with $(\mu/\omega)^{1/2}$ as a radial
coordinate. The decline of both the upper and lower envelopes are
stunningly linear.  This reflects the small-amplitude nature of the
oscillations which are now nearly harmonic so that our previous
argument works better here than in the previous section.

\subsection{Transition between different oscillation modes}

The lepton asymmetry of the neutrino flux is crucial for
understanding a realistic supernova case. To develop a first
understanding of this situation we consider an asymmetric ensemble
with the initial condition $\bar{\bf P}(0)=\alpha {\bf P}(0)$ where
$0\leq\alpha\leq 1$. If the $\nu$-$\nu$ interaction is sufficiently
strong, the two polarisation vectors will hang together by their
``internal magnetic field'' and all vectors ${\bf P}$, $\bar{\bf
P}$, ${\bf S}$, and ${\bf D}$ precess around ${\bf B}$ with the same
synchronised frequency. Using $\dot{\bf D}=\omega{\bf B}\times{\bf
S}$ and ${\bf D}={\bf S}(1-\alpha)/(1+\alpha)$, we thus find the
usual result
\begin{equation}\label{eq:omegasynch}
\omega_{\rm synch}=\frac{1+\alpha}{1-\alpha}\,\omega\,.
\end{equation}
Observe that the synchronised motion is faster for more symmetric
systems!

To achieve synchronised oscillations,  the internal frequencies of
the system, i.e., their precession frequency around a common spin,
must exceed the slow precession around the external magnetic field
$\omega_{\rm synch}$. In our case of two polarisation vectors, the
internal motion is described by $\dot{\bf S}=\mu{\bf D}\times{\bf
S}$. When synchronisation prevails, ${\bf D}$ has a fixed length
$(1-\alpha)$ so that the internal frequency is $(1-\alpha)\,\mu$.
Successful synchronisation means that this frequency must exceed
$\omega_{\rm synch}$, or, equivalently,
\begin{equation}\label{eq:synchsimple}
\frac{1+\alpha}{1-\alpha}\,\omega\alt(1-\alpha)\,\mu\,.
\end{equation}
This is exactly the argument and result presented in
Ref.~\cite{Duan:2005cp}. When this condition is violated, the motion
becomes bipolar.

Actually, this condition derives from an even simpler argument if we
consider the total energy of the system,
\begin{equation}
 H=\omega{\bf B}\cdot({\bf P}+\bar{\bf P})
 +\frac{\mu}{2}\,({\bf P}-\bar{\bf P})^2\,.
\end{equation}
Synchronised oscillations require that the energy of the system be
dominated by the spin-spin interaction, the second term. Assuming a
small vacuum mixing angle so that ${\bf B}$ is nearly aligned with
the initial polarisation vectors, and observing that $|{\bf
P}+\bar{\bf P}|=1+\alpha$ and $|{\bf P}-\bar{\bf P}|=1-\alpha$, the
requirement that the spin-spin term dominates is
$\omega\,(1+\alpha)\alt(1-\alpha)^2\,\mu/2$, identical with
Eq.~(\ref{eq:synchsimple}) up to a multiplicative factor.

In other words, synchronised oscillations occur when the
neutrino-neutrino part of the Hamiltonian always dominates over the
vacuum-oscillation part, even for the most disadvantageous
orientation of the polarisation vectors. At the other extreme, no
collective effects obtain when the neutrino-neutrino part never
dominates, even for the most advantageous orientation of these
vectors, i.e., when $\mu\alt\omega$. The bipolar regime corresponds
to the intermediate range where the relative magnitude of the
different energy contributions depends on the orientation of the
polarisation vectors. In summary, bipolar oscillations are expected
when
\begin{equation}
\omega\alt \mu<4 \frac{(1+\alpha)}{(1-\alpha)^2}\,\omega,
\end{equation}
and thus occur for an intermediate strength of the neutrino-neutrino
interaction, i.e., an intermediate range of neutrino densities as
first discussed in Ref.~\cite{Duan:2005cp}. The exact numerical factor
on the r.h.s.\ of this equation is taken from
Eq.~(\ref{eq:bipolcondition}) below.

\subsection{Asymmetric system as a pendulum with spin}

This simple reasoning gives us the correct scale for the
transition between synchronised and bipolar oscillations, but does
not explain the nature of the transition. The two oscillation modes
must be extreme cases of a continuum, yet the nature of the
intermediate cases is not obvious. It turns out that this continuum
has a straightforward physical interpretation that is quite
illuminating for an understanding of the entire system.

To this end we note that the asymmetric system is described by the
same equations of motion (\ref{eq:eom5}) for ${\bf Q}$ and ${\bf D}$
as the symmetric case, and $Q=|{\bf Q}|$ is likewise conserved. The
new feature here is the initial condition ${\bf D}(0)\not=0$ and
${\bf D}\cdot{\bf Q}\not=0$, so that Eq.~(\ref{eq:eom5}) now implies
\begin{equation}\label{eq:asymmetricd}
{\bf D}=\frac{1}{\mu}\,\frac{{\bf Q}\times\dot{\bf Q}}{Q^2}
+\frac{{\bf D}\cdot{\bf Q}}{Q^2}\,{\bf Q}\,.
\end{equation}
Since
\begin{equation}
\sigma={\bf D}\cdot{\bf Q}/Q
\end{equation}
is a constant of the motion, the expression~(\ref{eq:asymmetricd}) can
be equivalently written as
\begin{equation}
{\bf D}=\frac{{\bf q}\times\dot{\bf q}}{\mu}+\sigma\, {\bf q}\,,
\end{equation}
with ${\bf q}={\bf Q}/Q$.  The first term in the expression
corresponds to the orbital angular momentum as before, while the
second term plays the role of an inner angular momentum (i.e., spin)
of the pendulum's bob. This spin is always along the direction ${\bf
q}$, implying that we should think of the system as a spinning top
mounted in a way that its axis of rotation can swing like a
pendulum. If the top has no spin (i.e., $\sigma=0$), it acts as an
ordinary spherical pendulum.

Starting with the equation of motion $\dot{\bf D}=\omega{\bf
B}\times{\bf Q}$ and observing that $\dot{\bf q}\times\dot{\bf q}=0$,
we find
\begin{equation}
\frac{{\bf q}\times\ddot{\bf q}}{\mu}
+\sigma\, \dot{\bf q}=\omega Q{\bf B}\times{\bf q}\,.
\end{equation}
For the case $\sigma=0$, we recover the equation of motion of a
pendulum swinging in the plane defined by ${\bf B}$ and the initial
vector ${\bf q}(0)$ which we always choose to be in the
$z$-direction. Conversely, if $\mu$ is so large that we can neglect
the first term, we are back to a spin-precession equation with a
precession frequency $\omega_{\rm precess}=\omega Q/\sigma$. For
$\mu\gg\omega$ we have
\begin{eqnarray}\label{eq:alphameaning}
 Q&\approx&1+\alpha\,,\nonumber\\
 \sigma&\approx&1-\alpha\,,
\end{eqnarray}
so $\omega_{\rm precess}$ is indeed equal to $\omega_{\rm synch}$ of
Eq.~(\ref{eq:omegasynch}).

The extreme cases thus have an intuitive interpretation.  Synchronised
oscillations correspond to the flavour top spinning so fast that its
response to the force field is precession, just like a spun-up top on
a flat table surface.  On the other hand, if the top spins slowly
(corresponding to the case of a small neutrino asymmetry), then it
swings like an ordinary pendulum; the inner angular momentum in this
case has little impact, and we are in the bipolar mode.

Let us consider the asymmetric system in more detail.  Several
conserved quantities are apparent.  One is the energy of the system,
\begin{eqnarray}\label{eq:topenergy}
E&=&E_{\rm pot}+E_{\rm kin}\nonumber\\
&=&\omega Q \, ({\bf B}\cdot{\bf q}+1)
+\frac{\mu}{2} {\bf D}^2\nonumber\\
&=&\omega Q\,({\bf B}\cdot{\bf q}+1)
 + \frac{1}{2\mu}\,\dot{\bf q}^2
+\frac{\mu}{2}\,\sigma^2\,,
\end{eqnarray}
where we have added a constant to the potential energy such that it
vanishes when the pendulum is oriented opposite to the force field.
The other conserved quantity is the projection of ${\bf D}$ in the
${\bf B}$ direction, corresponding to the conservation of that
component of the angular momentum parallel to the force field and
which is thus not subject to a torque.

The conservation of angular momentum implies that the initial lepton
asymmetry cannot be changed beyond the amount caused by ordinary
vacuum oscillations. In all practical cases we begin with ${\bf D}$
oriented along the $z$-axis, while ${\bf B}$ is tilted by $2\theta_0$.
If $\theta_0$ is small, the initial neutrino asymmetry is almost
perfectly conserved.  Thus the self-interacting system cannot
stimulate an exotically large flavour conversion effect.

If the spin is not quite fast enough for perfect precession, the
overall motion is a wobble. The spinning top starts in a nearly
upright position, its axis pointing nearly to the north pole.  It will
tilt a bit until it reaches a certain latitude, when its motion
reverses back to the north pole. The latitude of reversal will lie
further south if either or both of $\sigma$ and $\mu$ is smaller.  In
other words, a smaller $\nu$-$\nu$ term makes the system ``more
bipolar''.  A more symmetric system (smaller $\sigma$) is also more
bipolar.

The southernmost position the pendulum can reach is defined by
energy and angular momentum conservation. As an example, we take the
mixing angle to be very small so that ${\bf B}$ is very close to the
$z$-direction. In this approximation, the initial total energy is
$E=2\omega Q+\mu \sigma^2/2$ and energy conservation
implies
\begin{equation}\label{eq:energycon}
\omega Q(1-\cos\varphi)=\frac{\dot{\bf q}^2}{2\mu}\,.
\end{equation}
On the other hand, angular momentum conservation along the ${\bf
B}$-direction gives
\begin{equation}\label{eq:angmomcon}
\sigma=\sigma\cos\varphi+\mu^{-1}\dot q_{\perp}\sin\varphi\,,
\end{equation}
where $\dot q_\perp$ is the velocity perpendicular to ${\bf
B}$, i.e., the pendulum's velocity along a circle of latitude.  The
largest excursion $\varphi_{\rm max}$ is reached when the pendulum
reverses its motion at its southernmost position where $\dot{\bf
q}^2=\dot q_{\perp}^2$.  Combined with Eqs.~(\ref{eq:energycon}) and
(\ref{eq:angmomcon}), we find
\begin{equation}\label{eq:cosphimax}
\cos\varphi_{\rm max}=\frac{\mu\sigma^2}{2\omega Q} -1\,
\end{equation}
for the largest excursion angle.

As expected, if either or both of $\mu$ and $\sigma$ becomes smaller,
Eq.~(\ref{eq:cosphimax}) tells us that the pendulum
reaches more southern latitudes. On the other hand, the equation has
no solution for
\begin{equation}\label{eq:synchronised}
\mu\sigma^2>4\omega Q\,.
\end{equation}
This condition corresponds to the fully synchronised case, which
prohibits any deviation from perfect $z$-alignment because of our
artificial assumption of ${\bf B}$ and the initial $\bf{P}$ and
$\bar{\bf{P}}$ being exactly aligned. If the pendulum had not
initially been perfectly aligned with ${\bf B}$, solutions would exist
for all values of the parameters. Still, for small mixing angles,
Eq.~(\ref{eq:synchronised}) provides an excellent estimate of the
condition for synchronised behaviour.  Using
Eq.~(\ref{eq:alphameaning}), this condition is equivalent, in the
$\mu/\omega\gg 1$ limit, to
\begin{equation}\label{eq:bipolcondition}
\frac{(1-\alpha)^2}{1+\alpha}>4\,\frac{\omega}{\mu},
\end{equation}
where, we recall, $\alpha$ parameterises the lepton asymmetry by
virtue of $|\bar{\bf P}|=\alpha|{\bf P}|$.

Taking our previous asymmetric example with $P_z(0)=1$ and $\bar
P_z(0)=0.8$ (i.e., $\alpha=0.8$) and assuming $\mu\gg \omega$,
synchronised behaviour is expected for $\mu/\omega>180$ or
$(\mu/\omega)^{1/2}>13.4$. This estimate corresponds well with the
onset of synchronisation in the top panel of Fig.~\ref{fig:musurvival}
where the mixing angle is large. Of course, the true point of onset
also depends logarithmically on the mixing angle---see the other
panels of Fig.~\ref{fig:musurvival}.

It is now evident that the onset of bipolar oscillations does not
imply full conversions. As we move into the bipolar regime,
the spinning top begins to wobble, reaching only some southern
latitude, but not the south pole. How far south it will get, i.e., the
spread between the upper and lower envelopes in
Fig.~\ref{fig:musurvival}, depends on the details of how the system
enters the bipolar regime. If the mixing angle is large, bipolar
oscillations begin almost immediately so that the amplitude of the
oscillations will be small.  If the mixing angle is small, the delayed
onset of the first bipolar swing allows $\mu$ to decrease further,
thereby resulting in a more southern turning point. Therefore, smaller
mixing angles imply a later onset of oscillations and a larger spread
between the envelopes. This is borne out by the examples
shown in Fig.~\ref{fig:musurvival}, where $\sin2\tilde\theta_0=0.1$,
$10^{-3}$ and $10^{-5}$ from top to bottom.

\subsection{Equipartition of Energies}

\label{sec:equipartition}

We have noted that the energy of the spinning top decreases in
proportion to $\mu^{1/2}$ once it has entered the bipolar regime,
assuming the decline of $\mu$ is sufficiently slow. It is illuminating
to note that from that time onward, the total energy
Eq.~(\ref{eq:topenergy}) is equipartitioned between $E_{\rm pot}$,
the potential energy in the external force field, and
$E_{\rm kin}$,  the internal and orbital kinetic energy of
the spinning top  (equivalent to the neutrino-neutrino interaction
energy).

\begin{figure}[t]
\begin{center}
\epsfig{file=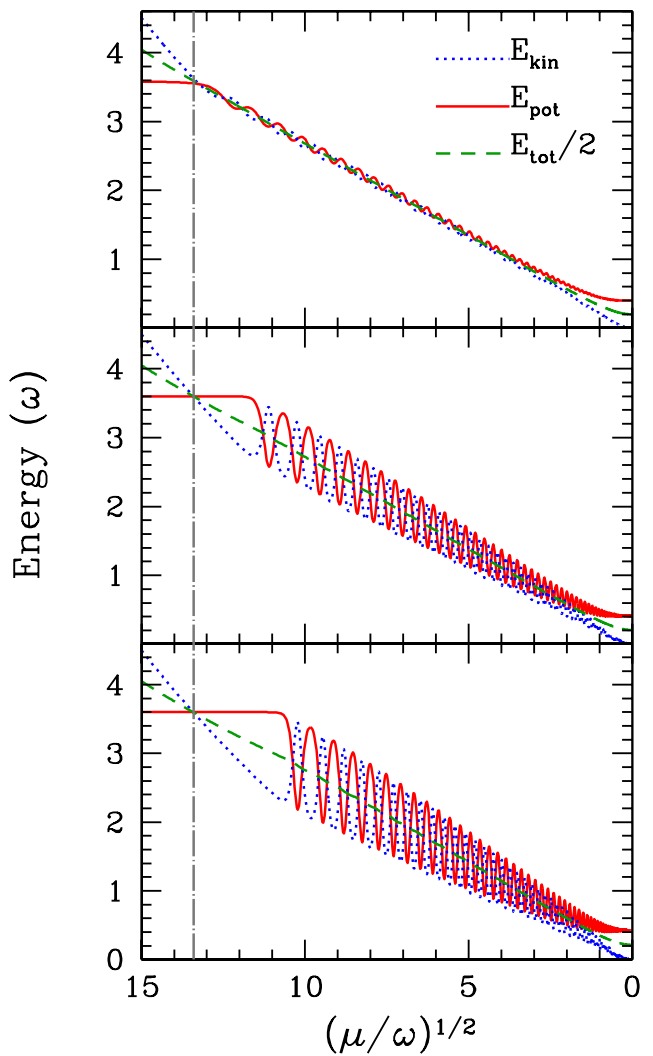,width=8.5cm}
\end{center}
\caption{Different energy components $E_{\rm pot}$, $E_{\rm kin}$ and
$E_{\rm tot}/2$ for the schematic supernova model of
Fig.~\ref{fig:SN2}, using $(\mu/\omega)^{1/2}$ as a radial
coordinate. The panels from top to bottom have the vacuum mixing angle
$\sin2\tilde\theta_0=0.1$, $10^{-3}$ and $10^{-5}$, respectively.  The
vertical line at $\mu/\omega=\sqrt{180}\approx 13.4$ marks the
transition to the bipolar regime according to
Eq.~(\ref{eq:bipolcondition}) for this example where
$\alpha=0.8$.}\label{fig:energies}
\end{figure}

To illustrate this point we show for our toy supernova the evolution
of $E_{\rm pot}$, $E_{\rm kin}$ and $E_{\rm tot}/2=(E_{\rm pot}+E_{\rm
kin})/2$ in Fig.~\ref{fig:energies}, using once more
$(\mu/\omega)^{1/2}$ as a radial coordinate.  Observe that indeed
$E_{\rm pot}=E_{\rm kin}$ at $\mu/\omega=180$ or
$(\mu/\omega)^{1/2}=13.4$ for our example $\alpha=0.8$.  It is
intriguing that the transition between the synchronised and the
bipolar regime is practically independent of the mixing angle. The
same is true also for the total energy $E_{\rm tot}$, which decreases
very nearly as $\mu^{1/2}$ in the bipolar regime as explained
earlier. In the synchronised regime close to the supernova, the
potential energy does not depend on $\mu$ whereas the kinetic energy
decreases with $\mu$.  In the bipolar region, on the other hand, the
kinetic and potential energies are nearly equipartitioned after
averaging over the pendulum's nutation period.

To understand this equipartition effect analytically, we consider a
case where the nutation amplitude is very small as in the top panel of
Fig.~\ref{fig:energies} (large mixing angle), so that it suffices to
study only the precession, i.e., we assume the pendulum's orbital
motion is such that its velocity is along a circle of latitude. As a
function of excursion angle $\varphi$, the total energy is
\begin{equation}
E_{\rm tot}=\omega Q(1+\cos\varphi)+\frac{\dot{\bf
q}^2}{2\mu}+\frac{\mu}{2}\sigma^2\,.
\end{equation}
Using angular-momentum conservation Eq.~(\ref{eq:angmomcon}) and
the relation $\dot{\bf q}^2=\dot q_{\perp}^2$ to eliminate the orbital
velocity, we find
\begin{equation}
 E_{\rm tot}=\omega Q(1+\cos\varphi)+ \frac{\mu\sigma^2}{2}\,
 \frac{1}{1+\cos\varphi}\,.
\end{equation}
We further recall that the system enters the bipolar regime when
$\mu\sigma^2=4\omega Q$, and that at this point the
excursion angle is still small so that $\cos\varphi=1$.
Therefore,  $E_{\rm pot}=E_{\rm kin}=2\omega Q$ at the onset of the
bipolar regime.  Subsequently,
$E_{\rm tot}$ is expected to scale as $\mu^{1/2}$ so that
 $E_{\rm tot}=(4\omega
Q\mu\sigma^2)^{1/2}$. We can now solve for the expression
$(1+\cos\varphi)$ as a function of $\mu$ and find explicitly $E_{\rm
pot}=E_{\rm kin}=(\omega Q\mu\sigma^2)^{1/2}$.

An important detail is that equipartition cannot hold all the way to
very small $\mu$. Angular momentum conservation, i.e., the approximate
conservation of the net $\nu_e$ flux in the limit of a small vacuum
mixing angle, implies that the potential energy is bounded from
below. In terms of polarisation vectors, this means that the strict
conservation of ${\bf B}\cdot{(\bf P}- \bar{\bf P})$ leads to an
approximately constant $P_z-\bar P_z=1-\alpha=\sigma$ in the case of
small vacuum mixing. Therefore, the smallest allowed value of the
potential energy is
\begin{equation}
\label{eq:Epot_min}
E_{\rm pot}\geq 2(1-\alpha)\omega=2\sigma\omega\,.
\end{equation}
In the example of Fig.~\ref{fig:energies} we have used an asymmetry
$\alpha=0.8$.  This gives an analytic estimate of $E_{\rm pot}\geq
0.4\,\omega$ using Eq.~(\ref{eq:Epot_min}), which is in excellent
agreement with numerical results. On the other hand, the kinetic
energy, being proportional to $\mu$, must eventually
vanish. Therefore, $E_{\rm pot}$ and $E_{\rm kin}$ approach different
limits as $\mu\to0$, as borne out by Fig.~\ref{fig:energies}.

Therefore, the evolution of the system now appears in a different
light. When $\mu$ decreases slowly, the system never properly enters
the bipolar regime: it stays at the edge of it. The polarisation
vectors spiral and slowly tilt in such a way that their energy in the
external $B$-field and the internal spin-spin energies stay equal to
the degree allowed by net lepton number conservation.

\section{Many Modes}
\label{sec:manymodes}

\subsection{Multiple frequencies}

\label{sec:muliplefrequencies}

We now turn to a more realistic case of an ensemble of $\nu_e$ and
$\bar\nu_e$ with different energies, i.e., different vacuum
oscillation frequencies $\omega_i$. The equations of motion are
\begin{eqnarray}\label{eq:many1}
 \partial_t{\bf P}_i&=&\left[+\omega_i{\bf B}
 +\mu\left({\bf P}-\bar{\bf P}\right)\right]\times{\bf P}_i\,,
 \nonumber\\
 \partial_t\bar{\bf P}_i&=&\left[-\omega_i{\bf B}
 +\mu\left({\bf P}-\bar{\bf P}\right)\right]\times\bar{\bf P}_i\,,
\end{eqnarray}
where for $N$ modes we use
\begin{equation}\label{eq:many2}
 {\bf P}=\sum_{i=1}^N{\bf P}_i\hbox{\qquad and\qquad}
 \bar{\bf P}=\sum_{i=1}^N\bar{\bf P}_i\,.
\end{equation}
We keep the normalisation $|{\bf P}|=|\bar{\bf P}|=1$ for the entire
ensemble so that the individual modes are normalised to $|{\bf
P}_i|=|\bar{\bf P}_i|=N^{-1}$.

In full analogy to the previous treatment we introduce the vectors
${\bf S}_i={\bf P}_i+\bar{\bf P}_i$ and ${\bf D}_i={\bf
P}_i-\bar{\bf P}_i$ as well as ${\bf D}=\sum {\bf D}_i ={\bf
P}-\bar{\bf P}$ so that
\begin{eqnarray}\label{eq:many3}
 \dot{\bf S}_i&=&\omega_i{\bf B}\times{\bf D}_i
 +\mu{\bf D}\times {\bf S}_i\,,
 \nonumber\\
 \dot{\bf D}_i&=&\omega_i{\bf B}\times{\bf S}_i
 +\mu{\bf D}\times {\bf D}_i\,.
\end{eqnarray}
Each pair of modes ${\bf P}_i$ and $\bar{\bf P}_i$ evolves
symmetrically so that each ${\bf D}_i$ is always oriented along the
$y$-axis and the terms ${\bf D}\times {\bf D}_i$ vanish.

We now assume strong coupling with $\mu/\omega_i \gg 1$ for all
modes so that we can also drop the $\omega_i{\bf B}\times {\bf D}_i$
term. This leaves us with the approximate equations of motion
\begin{eqnarray}\label{eq:many4}
 \dot{\bf S}_i&=&\mu{\bf D}\times {\bf S}_i\,,
 \nonumber\\
 \dot{\bf D}_i&=&\omega_i{\bf B}\times{\bf S}_i\,.
\end{eqnarray}
Equation~(\ref{eq:many4}) implies that all ${\bf S}_i$ evolve in the
same way because they precess in the same field ${\bf D}$. Flavour
conversion is now described by a single vector ${\bf S}=\sum {\bf
S}_i$ (and thus ${\bf S}_i={\bf S}/N$), and governed by
\begin{eqnarray}\label{eq:many5}
 \dot{\bf S}&=&\mu{\bf D}\times {\bf S}\,,
 \nonumber\\
 \dot{\bf D}&=&\left(\frac{1}{N}\sum_i^N
 \omega_i\right){\bf B}\times{\bf S}\,.
\end{eqnarray}
Thus, the evolution of the flavour content proceeds in the same way
as before [cf.\ Eq.~(\ref{eq:eom4})], but with the role of $\omega$
replaced with the average oscillation frequency of all modes
$\langle\omega\rangle\equiv N^{-1}\sum_i^N \omega_i$.

On the level of the individual modes, the ``tilting motion'' around
the $y$-axis is the same for all neutrino and antineutrino modes, so
in this sense their motion is synchronised. On the other hand, the
transverse motion characterised by ${\bf D}_i$ is different for
every mode because before summing, the equations of motion are
\begin{eqnarray}\label{eq:many6}
 \dot{\bf S}&=&\mu{\bf D}\times {\bf S}\,,
 \nonumber\\
 \dot{\bf D}_i&=&\omega_i\frac{{\bf B}\times{\bf S}}{N}\,,
\end{eqnarray}
where we have used ${\bf S}_i={\bf S}/N$.

\begin{figure}[t]
\begin{center}
\epsfig{file=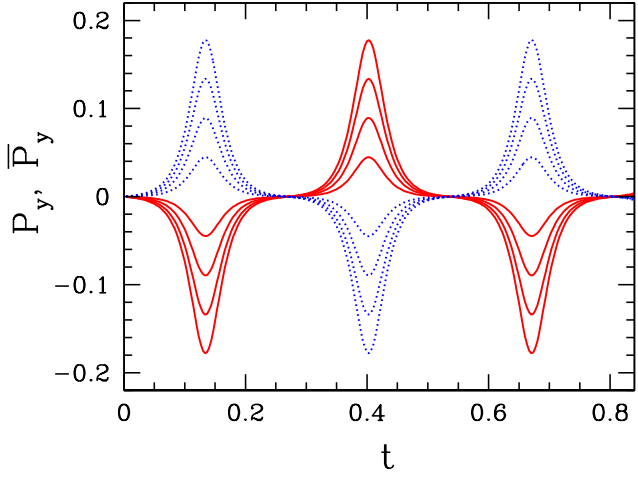,width=8.5cm}
\end{center}
\caption{Evolution of the individual modes $P_y$ (red/solid) and
$\bar{P}_y$ (blue/dotted) in a 4-mode system described by
Eq.~(\ref{eq:many1}), with $\tilde\theta_0=0.01$ (i.e., inverted
hierarchy), $\omega_i=1,2,3,4$ for $i=1$, 2, 3, and 4, respectively,
and strong neutrino-neutrino interaction
$\mu=100$.}\label{fig:pmany}
\end{figure}

It is, therefore, incorrect to say that all modes form two block spins
${\bf P}$ and $\bar{\bf P}$ which evolve separately in the bipolar
sense, with each individual mode staying aligned with its respective
block spin.  Separate alignment for neutrinos and antineutrinos was
explicitly claimed, for example, above Eq.~(4.1) in
Ref.~\cite{Kostelecky:1995dt}, in reference to the authors' own
numerical studies of Refs.~\cite{Kostelecky:1993dm,
Kostelecky:1994ys}. However, the observation of separate alignment
does not appear to be documented or demonstrated in these papers.
Whatever the origin of these authors' observation of bipolar
alignment, it is in conflict with our analytic treatment. We have
numerically verified that bipolar alignment does not hold and that our
individual ${\bf P}_i$ and $\bar{\bf P}_i$ vectors do indeed evolve
differently, as shown in Fig.~\ref{fig:pmany}.

The $z$-components of all modes evolve identically while the
transverse motion is different for modes with different $\omega_i$.
The transverse motion for the neutrino and antineutrino polarisation
vectors are opposite so that neutrinos and antineutrinos form two
distinct cohorts. In this sense the evolution actually {\it is}\/
bipolar. Therefore, we stick to this established terminology,
keeping in mind a broad interpretation of the word ``bipolar.''

The most important conclusion is that the strongly interacting
multi-mode system is exactly equivalent to one mode if one
interprets the vacuum oscillation frequency $\omega$ as the average
of all modes. Therefore, the entire system is still characterised by
a single collective variable. This conclusion also holds
for the asymmetric system of unequal densities of $\nu$ and
$\bar\nu$. The variation of different vacuum oscillation frequencies
is not a source of kinematical decoherences and the system behaves,
in this sense, similarly to synchronised oscillations.

\subsection{Different interaction strengths}

Instead of different frequencies one may also consider different
coupling constants between different modes.  In this case, the
general equations of motion are
\begin{eqnarray}\label{eq:eom3a}
 \partial_t{\bf P}_i&=&\left[+\omega_i{\bf B}
 +\sum_{j=1}^N\mu_{ij}\left({\bf P}_j-\bar{\bf P}_j\right)\right]
 \times{\bf P}_i\,,
 \nonumber\\
 \partial_t\bar{\bf P}_i&=&\left[-\omega_i{\bf B}
 +\sum_{j=1}^N\mu_{ij}\left({\bf P}_j-\bar{\bf P}_j\right)\right]
 \times\bar{\bf P}_i\,.
\end{eqnarray}
Variations in the interaction coefficients are motivated by their
dependence on the relative angle of the neutrino trajectories.
Neutrinos streaming off a supernova core are far from isotropic so
that unequal $\mu_{ij}$ coefficients are unavoidable. The recent
multi-angle simulations were aiming precisely at this
issue~\cite{Duan:2006an, Duan:2006jv}.

Even in an isotropic medium the coupling constants between the
different modes are never equal, but involve a factor
$(1-\cos\theta_{\bf pq})$ from the current-current structure of the
weak-interaction Hamiltonian. However, isotropy implies that all
modes with different momenta ${\bf p}$ but identical $p=|{\bf p}|$
evolve in the same way. Therefore, the angular part of the integral
in Eq.~(\ref{eq:eom1}) can be trivially performed in the sense that
the $\cos\theta_{\bf pq}$ term averages to zero. The isotropic
system is thus equivalent to the case of multiple frequencies, but a
common coupling constant described in
Sec.~\ref{sec:muliplefrequencies}. In other words it is equivalent
to the ``single-angle''~case.

In terms of our variables ${\bf S}_i$ (the flavour-dependent
particle plus antiparticle number) and ${\bf D}_i$ (the net lepton
number), the equations of motion are
\begin{eqnarray}\label{eq:eom3b}
 \dot{\bf S}_i&=&\omega_i{\bf B}\times{\bf D}_i
 +\sum_{j=1}^N\mu_{ij}{\bf D}_j\times{\bf S}_i\,,
 \nonumber\\
 \dot{\bf D}_i&=&\omega_i{\bf B}\times{\bf S}_i
 +\sum_{j=1}^N\mu_{ij}{\bf D}_j\times{\bf D}_i\,.
\end{eqnarray}
For the global ${\bf S}$ and ${\bf D}$ vectors this implies
\begin{eqnarray}\label{eq:eom3c}
 \dot{\bf S}&=&\sum_{i=1}^N\omega_i{\bf B}\times{\bf D}_i
 +\sum_{i,j=1}^N\mu_{ij}{\bf D}_j\times{\bf S}_i\,,
 \nonumber\\
 \dot{\bf D}&=&\sum_{i=1}^N\omega_i{\bf B}\times{\bf S}_i\,,
\end{eqnarray}
where we have used the symmetry $\mu_{ij}=\mu_{ji}$.  Equation
(\ref{eq:eom3c}) cannot be brought into the form of a closed set of
equations. However, even in this general case the quantity ${\bf
B}\cdot{\bf D}$ is a constant of the motion. This means that bipolar
oscillations never lead to lepton number conversions beyond the
amount caused by vacuum mixing.

It is possible to formulate a sufficient condition for a
multiple-coupling constant system to behave as a simple flavour
pendulum. Starting with Eq.~(\ref{eq:eom3b}) we note that the system
acts as a single flavour pendulum if all ${\bf S}_i$ and ${\bf D}_i$
tilt with the same speed. Self-consistency then requires that all
frequencies are equal, $\omega_i\equiv\omega$, and that
\begin{equation}\label{eq:musum}
\sum_{j=1}^N \mu_{ij}=\mu\,,
\end{equation}
with $\mu$ a number that is independent of $i$. In addition,
symmetry between different modes requires that $\mu_{ij}=\mu_{ji}$.
These conditions are met, for example, if
\begin{equation}
\mu_{ij}\propto g\left(\frac{i-j}{N}\right)\,,
\end{equation}
where $g(x)$ is an even function that is periodic in the sense
$g(x+1)=g(x)$. An example is $g(x)=\cos(2\pi x)$.

An important case where these conditions are violated is neutrinos
radiated from a supernova core. Assuming overall spherical symmetry,
the only parameter that differentiates between trajectories is
$\cos\theta$ with $\theta$ the angle relative to the radial
direction. Considering instead the schematic case of neutrinos emitted
``isotropically'' by a plane surface, the coupling constants
are~\cite{Duan:2006an}, following Eq.~(\ref{eq:cicj}),
\begin{eqnarray}\label{eq:multiangle}
\mu_{ij}&=&\mu\,\frac{4}{3}\,(1-c_ic_j)\nonumber\\
 &=&\mu\,\frac{4}{3}\,
 \left[1-\frac{(i-\frac{1}{2})(j-\frac{1}{2})}{N^2}\right]\,.
\end{eqnarray}
Here, $c_i=\cos\theta_i$ is the cosine of the neutrino mode relative
to the radiating surface's normal direction. We assume a uniform
distribution in $0\leq c_i\leq 1$ represented by discrete modes as in
the second line of Eq.~(\ref{eq:multiangle}). Note that
$\sum_{i=1}^N(i-\frac{1}{2})=N^2/2$ so that the average
$\langle\mu_{ij}\rangle=\mu$ is exact, even for a small number of
modes.

This example does not satisfy the condition Eq.~(\ref{eq:musum})
because
\begin{equation}
 \sum_{j=1}^N \mu_{ij}=\mu\,\frac{4N}{3}
 \left[1-\frac{i-\frac{1}{2}}{2N}\right]\,.
\end{equation}
(Note that the overall factor of $N$ is compensated by using
individual polarisation vectors that are normalised to the length
$N^{-1}$.) Therefore, one cannot expect neutrinos streaming off a
supernova core to oscillate in a collective manner. Rather, one should
expect kinematical decoherence within a few bipolar periods,
i.e., on a time scale of order $\kappa^{-1}=(\omega\mu)^{1/2}$. In
other words, the length of the total ${\bf P}$ or $\bar{\bf P}$ vector
is no longer conserved and the ensemble partly or fully decoheres.

In simple numerical examples of a symmetric system this expectation
is indeed borne out, i.e., a non-isotropic ensemble consisting of
equal numbers of neutrinos and antineutrinos turns into an equal
mixture of both flavours within a few bipolar oscillation periods for
both the normal and inverted hierarchy.

On the other hand, the large-scale numerical studies of
Ref.~\cite{Duan:2006an} show that neutrinos streaming off a supernova
core are sometimes quite well represented by the single-angle case,
i.e., collective behaviour rather than quick kinematical decoherence
appears to be more generic. Again, the main differences to a simple
symmetric system are twofold: there is a neutrino-antineutrino
asymmetry and the effective density declines with radius.  We have
already observed in Sec.~\ref{sec:equipartition} that for the inverted
hierarchy the asymmetry, together with the slow decline of the
neutrino density, has the effect of slowly turning the polarisation
vectors without the system ever entering the bipolar regime, i.e., the
system teeters along the edge of the bipolar condition. Since the
bipolar regime is never properly entered, it is less surprising that
kinematical decoherence is not a prominent feature of the evolution.

A dedicated research project is required to develop a deeper
understanding of the conditions that determine if the system evolves
as as single collective system in the form of the flavour pendulum, or
if it kinematically decoheres in that the individual polarisation
vectors of the different modes are ``randomised'' in flavour space.

Our main conclusion is that different oscillation frequencies are not
a source of kinematical decoherence, while the multi-angle nature of
a non-isotropic system is such a source, especially for a symmetric
system. The detailed interplay between the collective mode represented
by our pendulum and the multi-modal nature of the non-isotropic system
remains to be understood.

\section{Connection to quantum physics}
\label{sec:quantum}

\subsection{Quantisation of the flavour pendulum}

We have seen that the flavour conversion of neutrinos streaming off a
supernova core can be understood almost completely when we cast the
equations of motion in the form of a pendulum in flavour space that
may include inner angular momentum if we need to account for a lepton
asymmetry. One may then ask if flavour conversion was
possible in the absence of flavour mixing since, even if the
pendulum were placed exactly on its tip, Heisenberg's uncertainty
relation would prevent it from staying there forever, just as an
idealised pencil cannot stand on its tip indefinitely.

To estimate the relevant time scale we recall that the equations of
motion for the pendulum's excursion angle can be derived from the
classical Hamiltonian Eq.~(\ref{eq:ham1}) with $\varphi$ and $D$ as
the canonical variables. In order to quantise this system, however, we
need to be more careful about absolute scales, and the equations of
motion for the quantum variables must follow from the Hamiltonian of
the full quantum system. In the case of $N$ neutrinos and $N$
antineutrinos, the full Hamiltonian is simply $N$ times the one of
Eq.~(\ref{eq:ham1}),
\begin{equation}\label{eq:ham1full}
H_{\rm tot}=\frac{N}{2}\,\mu\,D^2
+\frac{N\kappa^2}{\mu}\bigl[1-\cos(\varphi+2\theta_0)\bigr]\,.
\end{equation}
Identifying $\varphi$ as the canonical coordinate, the familiar
equations of motion (\ref{eq:eom6}) follow classically using
Hamilton's equations, provided we interpret $ND$ as the conjugate
momentum. The pendulum's potential energy in the macroscopic sense
scales with $N$, and likewise its moment of inertia $I=N/\mu$.  On the
quantum level, the corresponding commutation relation~is
\begin{equation}
[\varphi,D]=\frac{{\rm i}\hbar}{N}\,.
\end{equation}
Using this commutation relation, the same equations of motion
(\ref{eq:eom6}) follow quantum mechanically from the full
Hamiltonian~(\ref{eq:ham1full}) by way of Heisenberg's equations of
motion ${\rm i}\hbar \dot\varphi=[\varphi,H_{\rm tot}]$ and ${\rm
i}\hbar\dot D=[D,H_{\rm tot}]$.

The fact that the same equations of motion for the tilt angle
$\varphi$ follow both classically and quantum mechanically
irrespective of the size of $N$---provided we identify the appropriate
canonical momentum---indicates that the flavour evolution does not
depend on the size of the system.  Thus, as long as our calculation is
classical, we can work with polarisation vectors and associated
angular momenta that are normalised to unity.  On the quantum level,
however, the absolute length of the polarisation vectors will affect
the quantisation of the system, and it is necessary that we use the
correct Hamiltonian with the appropriate factors of $N$.

To estimate the time scale for the pendulum to stay upright we
consider its downward vertical position, i.e., we consider it to be a
harmonic oscillator in its quantum-mechanical ground state.  The
uncertainties of the canonical variables in this state are
\begin{equation}
\langle\varphi^2\rangle=\frac{1}{2}\,\frac{1}{I\kappa}\,,
\quad\hbox{and}\quad
\langle (ND)^2\rangle=\frac{1}{2}\,I\kappa\,,
\end{equation}
where $\kappa$ is the oscillation frequency of the pendulum and
$I=N/\mu$ its moment of inertia.  In the strong-interaction limit
$\mu\gg\omega$, the expression for $\langle\varphi^2\rangle$ becomes
\begin{equation}
 \langle \varphi^2\rangle=\frac{1}{2N}\,
 \left(\frac{\mu}{2\omega}\right)^{1/2}\,.
\end{equation}
Let us now put in some realistic numbers.  The typical density of
neutrinos in the supernova region of interest is estimated to be
$10^{32}~{\rm cm}^{-3}$ in Appendix \ref{sec:eom} before
Eq.~(\ref{eq:musn}).  The volume of the critical region may be of
order $100~{\rm km}^3$ so that some $10^{53}$ particles may be in the
system at any one time. Moreover, the bipolar regime begins at
$(\mu/\omega)^{1/2}$ of order 10.  Thus a typical value for
$\langle\varphi^2\rangle^{1/2}$ would be of order $10^{-26}$. If this
number is taken to be a typical excursion angle caused by quantum
fluctuations, then the time scale to tilt will be of order
$\kappa^{-1}\ln(10^{26})\approx 60 \ \kappa^{-1}$. Since $\kappa^{-1}$
is a fraction of a km, the quantum effect would happen over a length
scale of tens of~km. Of course, the vacuum mixing angle and thus the
initial excursion of the pendulum is much larger than this quantum
estimate. Moreover, the system would be subject to other
forces. Still, this estimate demonstrates that in an unstable system
exponential growth can quickly enhance quantum effects to a
macroscopic scale.

\subsection{Full quantum system}

The discussion above shows that our classical treatment of the flavour
evolution of a large neutrino ensemble was justified. However, it is
still instructive to briefly explain the structure of the full quantum
Hamiltonian. It is well known that the neutrino interaction in flavour
space has an $SU(2)$ structure and as such is equivalent to a spin
system~\cite{Friedland:2003eh, Friedland:2006ke,
Balantekin:2006tg}. The equations of motion with a matter term
(\ref{eq:matter1}) can be shown to follow from the quantum Hamiltonian
\begin{eqnarray}
\hat H&=&\hat H_0+\hat H_{\rm matter}+\hat H_{\nu\nu} \\
&=&\omega {\bf B}\cdot(\hat{\bf S}-\hat{\bar{\bf S}})
+\lambda{\bf L}\cdot(\hat{\bf S}+\hat{\bar{\bf S}})
+\frac{\mu}{N}(\hat{\bf S}+\hat{\bar{\bf S}})^2\,,\nonumber
\end{eqnarray}
where we now use carets to denote quantum operators explicitly.  The
operators $\hat{\bf S}$ and $\hat{\bar{\bf S}}$ each represent an
angular momentum $N/2$, i.e., $N$ spin $\frac{1}{2}$ particles that
are linked to form one big ``block spin'' each. The equation of motion
for $\hat{\bf S}$ follows from ${\rm i}\hbar\partial_t\hat{\bf S}
=[\hat{\bf S},\hat H]$ and the angular momentum commutation relation
$[\hat{S}_i,\hat{S}_j]={\rm i}\hbar\epsilon_{ijk}\hat{S}_k$ and
similarly for $\hat{\bar{\bf S}}$.  Note that $\hbar$ drops out of the
spin-precession equation $\partial_t\hat{\bf S}=\omega{\bf
B}\times\hat{\bf S}$. Therefore, spin precession is fundamentally a
classical phenomenon and one does not need to distinguish carefully
between equations of motion for quantum operators and for expectation
values. However, for the nonlinear neutrino-neutrino term, it is not
intuitively obvious that one can ignore correlation effects when
taking the expectation values~\cite{Friedland:2003dv,Bell:2003mg}.

The connection to the polarisation vectors is that ${\bf
P}=\langle\hat{\bf S}\rangle\,(2/N)$ is the normalised expectation
value of the spin that represents the particles, whereas $\bar{\bf
P}=-\langle\hat{\bar{\bf S}}\rangle\,(2/N)$ includes a minus sign. In
other words, the quantities ${\bf P}$ and $\bar{\bf P}$ play the role
of ``magnetic moments'' in flavour space, whereas the quantities
$\hat{\bf S}$ and $\hat{\bar{\bf S}}$ play the role of angular
momenta. We call them ``flavour spins,'' but the terminology
``neutrino flavour isospins (NFIS)'' has also been
used~\cite{Duan:2005cp, Duan:2006an, Duan:2006jv}.  The negative sign
between the polarisation vector and flavour spin for antineutrinos is
consistent with antiparticles of equal spin carrying negative magnetic
moments relative to the particles, such as the case of electrons and
positrons. This negative sign also explains that under the mass
Hamiltonian in vacuum $\hat{H}_0$, neutrinos and antineutrinos precess
``in opposite directions'' in flavour space. We find that the language
of polarisation vectors is useful in the classical limit, whereas the
language of flavour spins is useful when dealing with the quantum
aspects of the system.

We note that only the vacuum Hamiltonian $\hat H_0$ distinguishes
between neutrinos and antineutrinos, while the matter and $\nu$-$\nu$
parts of the full Hamiltonian can be expressed in terms of a single
big angular momentum operator $\hat{\bf J}=\hat{\bf S}+\hat{\bar{\bf
S}}$. Therefore, these parts of the Hamiltonian are equivalent for our
system consisting of neutrinos and antineutrinos, and a neutrino-only
system consisting of two flavours. In the neutrino-only case,
``flavour spin up'' means $\nu_e$, ``flavour spin down'' $\nu_\mu$. In
our case, ``flavour spin up'' means either $\nu_e$ or $\bar\nu_\mu$,
and ``flavour spin down'' $\bar\nu_e$ or $\nu_\mu$, and the states
$\nu_e$ and $\bar\nu_\mu$ are fully equivalent in the absence of $\hat
H_0$ (and similarly for $\bar\nu_e$ and $\nu_\mu$).

In terms of the big angular momentum operator $\hat{\bf J}$, we see
that the neutrino-neutrino Hamiltonian is of the form $\hat {\bf
J}^2$, while the interaction with ordinary matter is proportional to
$\hat J_z$. These two operators commute so that they have a common set
of energy eigenstates. This observation is the quantum analogue to our
classical result that the presence of ordinary matter leaves bipolar
oscillations nearly unaffected.

Some time ago it was speculated that a system of many spins
interacting by a nonlinear Hamiltonian of the form $\hat{\bf J}^2$
could exhibit quantum entanglement effects in the sense that its
evolution is coherently accelerated~\cite{Bell:2003mg}. Applied to our
case, this conjecture means the following. Consider a ``dense gas''
consisting of exactly one $\nu_e$ and one $\bar\nu_e$ with the same
density as our macroscopic system, i.e., with the same spin-spin
interaction energy $\mu$. In this case the four possible states of the
system are grouped into a triplet state consisting of
$\left|\nu_e,\bar\nu_\mu\right\rangle$,
$\frac{1}{\sqrt2}(\left|\nu_e,\bar\nu_e\right\rangle
+\left|\nu_\mu,\bar\nu_\mu\right\rangle)$ and
$\left|\nu_\mu,\bar\nu_e\right\rangle$, and a singlet state
$\frac{1}{\sqrt2}(\left|\nu_e,\bar\nu_e\right\rangle
-\left|\nu_\mu,\bar\nu_\mu\right\rangle)$. Put another way, the energy
eigenstates of the system are the usual angular momentum states
$\left|J,m\right\rangle$, where $J=0,1$ and $m=-J,-J+1,\ldots,J$, that
carry no ``magnetic moment.''  This is perfectly analogous to
positronium that consists of two spin-$\frac{1}{2}$ particles with
opposite magnetic moments. Neither the singlet nor the triplet state
of positronium carries a magnetic moment. However, we can prepare the
system in a state with magnetic moment, in our case a state like
$\left|\nu_e,\bar\nu_e\right\rangle= \frac{1}{\sqrt2}(
\left|1,0\right\rangle+\left|0,0\right\rangle)$. Because the energies
of the singlet and triplet states are split by the amount $\mu$, we
will obtain oscillations between the
$\left|\nu_e,\bar\nu_e\right\rangle$ and
$\left|\nu_\mu,\bar\nu_\mu\right\rangle$ states with frequency $\mu$.

Next, we make the system larger (more extensive) without changing its
intensive properties, i.e., we keep $\mu=\sqrt2 G_{\rm F} N/V$ fixed
with $N$ the number of neutrino-antineutrino pairs and $V$ the
volume. If the system is prepared in a state consisting of $N$
$\nu_e$-$\bar\nu_e$ pairs, will it convert to a state of
$\nu_\mu$-$\bar\nu_\mu$ pairs on a similar time scale $\mu^{-1}$?
That is, will the magnetic moment of a ``super-positronium''
consisting of $N$ electrons and $N$ positrons reverse on the same time
scale as for ordinary positronium? Formally, this amounts to solving
for the expectation value $\langle S_z- \bar{S}_z \rangle$ as a
function of time, and, using the results of
Refs.~\cite{Friedland:2003eh, Friedland:2006ke}, it can be shown that
the conversion time scale is of order $\sqrt{N}\mu^{-1}$, not
$\mu^{-1}$, i.e., much longer for a macroscopic ensemble.%
\footnote{References~\cite{Friedland:2003eh, Friedland:2006ke}
consider a neutrino-only system, initially prepared with $N$ $\nu_e$
and $M$ $\nu_x$, where $\nu_x$ is some linear combination of $\nu_e$
and $\nu_\mu$. However, because of the exact correspondence between
this system and our neutrino-antineutrino system (in the absence of
$H_0$) discussed earlier, the results of Refs.~\cite{Friedland:2003eh,
Friedland:2006ke} can be trivially mapped to our case.  In particular,
the connection between our $\langle S_z- \bar{S}_z \rangle$ and their
$P_1(t)$ can be found in Sec.~IIID of Ref.~\cite{Friedland:2006ke}.}

Therefore, quantum effects enter on the usual level of $1/\sqrt N$
fluctuations and do not cause any novel effects on a macroscopic
scale. Put another way, the equilibration in flavour space of a large
$\nu_e$-$\bar\nu_e$ ensemble requires a time scale corresponding to
ordinary pair processes
$\nu_e\bar\nu_e\leftrightarrow\nu_\mu\bar\nu_\mu$ that are of second
order in $G_{\rm F}$. To first order in $G_{\rm F}$, the flavour
equilibration requires vacuum mixing and the phenomenon of bipolar
oscillations.

\section{Discussion and Summary}
\label{sec:discussion}

We have studied bipolar neutrino oscillations, i.e., the flavour
evolution of an ensemble initially consisting of equal numbers of
$\nu_e$ and $\bar\nu_e$. We have shown that the classical equations of
motion can be cast in the form of a pendulum in flavour space. The
surprising bipolar conversion effect observed for the inverted mass
hierarchy corresponds to the pendulum starting in a near upright
position, with the excursion angle growing exponentially until the
pendulum makes an almost complete swing.  Conversely, if it starts in
a nearly vertical downward position (i.e., normal mass hierarchy), the
system behaves as a harmonic oscillator.

For the inverted case, the time for a complete swing is given by the
pendulum's oscillation period times a factor depending on the
logarithm of the initial excursion angle which is nearly identical
with twice the vacuum mixing angle. Therefore, the bipolar conversion
is delayed by the logarithm of the small vacuum mixing
angle. Likewise, we derived an analytic solution for the case when
ordinary matter is present and showed that it affects the bipolar
conversion time also only logarithmically.

If the vacuum oscillation frequencies are different for different
modes, one cannot represent the ensemble by two ``block spins.''
However, in a dense $\nu$-$\bar\nu$ gas our model remains unaffected
except that the vacuum oscillation frequency is replaced by an average
over all modes. When the coupling strength between different modes
varies, as would be realistically expected in a non-isotropic
ensemble, yet other forms of behaviour appear. In particular, the
different modes can kinematically decohere in flavour~space.  Indeed,
our results suggest---and it has been numerically observed in
simulations~\cite{Duan:2006an,Duan:2006jv}---that partial flavour
decoherence, instead of a simple swapping of flavours, is a possible
feature in a multi-coupling/multi-angle system.

However, the same simulations also suggest that the flavour evolution
of neutrinos streaming off a supernova core is qualitatively
approximated by a single-angle treatment, and that collective
behaviour appears to be the more generic outcome.  A partial
explanation for this, at least in the inverted hierarchy case, is
provided by our observation that the single-angle system with a
decreasing neutrino density and a non-zero neutrino-antineutrino
asymmetry never becomes properly bipolar. Rather, it evolves such that
the potential and kinetic energy of the pendulum remain
equipartitioned, signifying that the system remains at the edge of the
bipolar condition throughout its evolution and is hence better immuned
to decoherence.

The only apparent case of practical interest for this discussion is
flavour conversion of neutrinos streaming off a supernova core where
collective flavour transformations play an important role.
Close to the neutrino sphere, the oscillations are
synchronised up to a few tens of kilometres, then they enter the
bipolar regime, and finally, beyond 100--200 km, ordinary
oscillations occur~\cite{Duan:2005cp, Duan:2006an, Duan:2006jv}.
Ordinary matter effects modify the oscillations in the usual way
both in the synchronised regime and far away where collective
effects are irrelevant, whereas in the intermediate regime
of bipolar oscillations ordinary matter has no significant impact.
This counter-intuitive situation was conjectured and numerically
observed in Refs.~\cite{Duan:2005cp, Duan:2006an, Duan:2006jv}.
In our model of a flavour pendulum the impact of ordinary matter
can be calculated analytically.

Note that while the bipolar behaviour extends over a large range in
radius outside the neutrino sphere, we have explicitly assumed in
our treatment that the system consists of only two flavours because
only one mass splitting is of importance. However, it could well be
that the solar mass difference $\Delta m_{\rm solar}^2 \sim 8 \times
10^{-5} \, {\rm eV}^2$ cannot be ignored in the whole region. If
this is the case a full three-flavour description must be employed,
and new phenomena might arise.

In any case, collective neutrino oscillations, unsuppressed by
ordinary matter, in the region a few tens of kilometres above the
neutrino sphere will likely change the picture of supernova flavour
oscillations and observable consequences in various ways.  Taking the
atmospheric mass hierarchy to be inverted, the ``single-angle
approximation'' predicts a swapping of the $\bar\nu_e$ and
$\bar\nu_\mu$ as well as of the $\nu_e$ and $\nu_\mu$ fluxes with a
possible impact on r-process nucleosynthesis~\cite{Pantaleone:1994ns,
Qian:1994wh, Sigl:1994hc, Qian:1993dg}, energy transfer to the
stalling shock wave~\cite{Fuller1992}, and the possibility to observe
shock-wave propagation effects in neutrinos~\cite{Schirato:2002tg,
Takahashi:2002yj, Lunardini:2003eh, Fogli:2003dw, Tomas:2004gr,
Dasgupta:2005wn, Fogli:2004ff, Choubey:2006aq, Fogli:2006xy}.  Nothing
new happens in the case of a normal mass hierarchy, so that one still
expects observable effects such as Earth matter effects in the
neutrino signal from the next galactic supernova~\cite{Dighe:1999bi,
Lunardini:2001pb, Dighe:2003be, Chiu:2006, Mirizzi:2006xx}. However,
this conclusion assumes the validity of the single-angle treatment.
Partial or complete kinematical decoherence, caused by the multi-modal
nature of the system, will affect the flavour composition of the
neutrinos passing the bipolar region even in the normal hierarchy.

The one case that probably remains unaffected is the
prompt deleptonisation burst where initially the $\nu_e$ flux is
strongly enhanced relative to $\nu_{\mu}$, $\bar\nu_{\mu}$,
$\nu_{\tau}$ and $\bar\nu_{\tau}$, while the $\bar\nu_e$ flux is
strongly suppressed~\cite{Kachelriess:2004ds}.  In this case, the
bipolar condition is not fulfilled and one expects ``ordinary''
synchronised oscillations.


\begin{acknowledgments}
We thank H.~Duan, S.~Pastor, M.~Sloth, and R.~Tom\`as for illuminating
discussions and acknowledge A.~Mirizzi, B.~Dasgupta, H.~Duan,
G.~Fuller, S.~Pastor for important comments on the manuscript. This
work was partly supported by the Deutsche Forschungsgemeinschaft under
Grant No.~SFB-375 and by the European Union under the Ilias project,
contract No.~RII3-CT-2004-506222. SH acknowledges support from the
Alexander von Humboldt Foundation through a Friedrich Wilhelm Bessel
Award.
\end{acknowledgments}

\bigskip

\appendix
\section{General equations of motion}
\label{sec:eom}

\subsection{Multiflavour system}

We summarise here the general equations of motion for the flavour
evolution of an ensemble of mixed neutrinos. Our main purpose is to
show the meaning of the different terms in the general context and
their relative signs and to establish our conventions.

A statistical ensemble of unmixed neutrinos is characterised by the
occupation numbers $f_{\bf p}=\langle a^\dagger_{\bf p} a_{\bf
p}\rangle$ for each momentum mode ${\bf p}$, where $a^\dagger_{\bf
p}$ and $a_{\bf p}$ are the relevant creation and annihilation
operators and $\langle\ldots\rangle$ is the expectation value.
A corresponding expression can be defined for the antineutrinos,
${\bar f}_{\bf p}=\langle {\bar a}^\dagger_{\bf p} {\bar a}_{\bf
p}\rangle$, where overbarred quantities always refer to
antiparticles. In a multiflavour system of mixed neutrinos, the
occupation numbers are generalised to density matrices in flavour
space~\cite{Dolgov:1980cq,Sigl:1992fn,mckellar&thomson}
\begin{eqnarray}\label{eq:densitymatrixdefinition}
 (\rho_{\bf p})_{ij}&=&
 \langle a^\dagger_{i} a_{j}\rangle_{\bf p}\,,\nonumber\\
 (\bar \rho_{\bf p})_{ij}&=&
 \langle \bar a^\dagger_{j}\,\bar a_{i}\rangle_{\bf p}\,.
\end{eqnarray}
The reversed order of the flavour indices $i$ and $j$ in the r.h.s.\
for antineutrinos is crucial to ensure that $\rho_{\bf p}$ and
$\bar\rho_{\bf p}$ behave consistently under a flavour
transformation. The seemingly intuitive equal order of flavour
indices that is frequently used in the
literature~\cite{Samuel:1993uw, Kostelecky:1994ys,
Kostelecky:1993yt, Kostelecky:1993dm, Kostelecky:1995dt,
Kostelecky:1995xc, Samuel:1996ri, Kostelecky:1996bs,
Pantaleone:1998xi, Duan:2005cp, Duan:2006an} causes havoc in that
$\rho_{\bf q}-\bar\rho_{\bf q}^*$ instead of $\rho_{\bf
q}-\bar\rho_{\bf q}$ appears in Eq.~(\ref{eq:eom1}). Therefore, the
equations of motion then involve $\rho_{\bf p}$, $\bar\rho_{\bf p}$,
$\rho^*_{\bf p}$ and $\bar\rho^*_{\bf p}$ and thus lose much of
their simplicity even if they are, of course, equivalent.

Flavour oscillations of an ensemble of neutrinos and antineutrinos
are described by
\begin{widetext}
\begin{eqnarray}\label{eq:eom1}
 \partial_t \rho_{\bf p}&=&-{\rm i}\left[\Omega_{\bf p}
 +\sqrt{2}\,G_{\rm F}\left(L-\bar L+
 \int\!\frac{d^3{\bf q}}{(2\pi)^3}
 \left(\rho_{\bf q}-\bar\rho_{\bf q}\right)(1-\cos\theta_{\bf pq})
 \right)
 ,\rho_{\bf p}\right],
 \nonumber\\
 \partial_t \bar\rho_{\bf p}&=&+{\rm i}\left[\Omega_{\bf p}
 -\sqrt{2}\,G_{\rm F}\left(L-\bar L+
 \int\!\frac{d^3{\bf q}}{(2\pi)^3}
 \left(\rho_{\bf q}-\bar\rho_{\bf q}\right)(1-\cos\theta_{\bf pq})
 \right)
 ,\bar\rho_{\bf p}\right],
\end{eqnarray}
\end{widetext}
where $[{\cdot},{\cdot}]$ is a commutator and $G_{\rm F}$ is the
Fermi constant. For ultrarelativistic neutrinos, the matrix of
vacuum oscillation frequencies is $\Omega_{\bf p}={\rm
diag}(m_1^2,m_2^2,m_3^2)/2p$ with $p=|{\bf p}|$ when expressed in
the mass basis. The ordinary matter effect is encapsulated in the
matrix of charged lepton densities, $L={\rm diag}(n_e,n_\mu,n_\tau)$
in the weak interaction basis, and in a corresponding matrix $\bar
L$ for the charged antilepton densities. The factor
$(1-\cos\theta_{\bf pq})$, where $\theta_{\bf pq}$ is the angle
between ${\bf p}$ and ${\bf q}$, will not average to unity if the
neutrino gas is not isotropic.

These and the more general Boltzmann kinetic equations apply only if
no correlations build up between the different
modes~\cite{Sigl:1992fn}. This condition may well be violated when
neutrino-neutrino interactions dominate~\cite{Bell:2003mg}, but does
not seem to be important in practice for ensembles of large numbers
of neutrinos~\cite{Friedland:2003dv, Friedland:2003eh,
Friedland:2006ke}.

\subsection{Two-flavour system}

Collective oscillation effects have been studied for the case of
two-flavour oscillations. The measured hierarchy of neutrino mass
differences suggests that oscillations driven by the ``atmospheric''
and ``solar'' mass differences occur at vastly different epochs in
the early universe and at vastly different distances from a
supernova core. Genuine three-flavour collective effects have not
been addressed in the literature.

The two-flavour system has the great advantage that all $2{\times}2$
matrices can be expressed in terms of the unit matrix and the Pauli
matrices with a vector of coefficients. Explicitly we write
\begin{eqnarray}\label{eq:matrices}
 \Omega_{\bf p}&=&{\textstyle\frac{1}{2}}
 \bigl(\Omega_0+\omega_{\bf p}\,{\bf B}
 \cdot\hbox{\boldmath$\sigma$}\bigr)\,,\nonumber\\
 L-\bar L&=&{\textstyle\frac{1}{2}}
 \bigl(n_0+n_e\,{\bf L}
 \cdot\hbox{\boldmath$\sigma$}\bigr)\,,\nonumber\\
 \rho_{\bf p}&=&{\textstyle\frac{1}{2}}
 \bigl(f_{\bf p}+{\bf P}_{\bf p}
 \cdot\hbox{\boldmath$\sigma$}\bigr)\,,\nonumber\\
 \bar\rho_{\bf p}&=&{\textstyle\frac{1}{2}}
 \bigl(\bar f_{\bf p}+\bar{\bf P}_{\bf p}
 \cdot\hbox{\boldmath$\sigma$}\bigr)\,.
\end{eqnarray}
The vectors ${\bf P}_{\bf p}$ and $\bar{\bf P}_{\bf p}$ are the
$\nu$ and $\bar\nu$ polarisation vectors in flavour space. We
choose the coordinate system in flavour space such that a polarisation
vector pointing in the positive $z$-direction signifies pure electron
neutrinos or antineutrinos, whereas an orientation in the negative
$z$-direction corresponds to muon neutrinos. In this convention the
polarisation vectors are not unit vectors in a flavour-pure system.
For example, in an ensemble of pure electron neutrinos, the
$z$-component of ${\bf P}_{\bf p}$ corresponds to the electron
neutrino occupation number, and the total number density of electron
neutrinos would be $n_{\nu_e}=\int P^z_{\bf p} \ d^3{\bf p}/(2\pi)^3$.
Of course, ${\bf P}_{\bf p}=0$ does not mean that this mode is empty;
it just means that it contains an incoherent equal mixture of electron
and muon neutrinos.

In an ordinary medium
there are no charged muons or tau-leptons. Therefore, ${\bf L}$ is a
unit vector in the positive $z$-direction and $n_e$ is an
effective electron density, i.e., the density of electrons minus
that of positrons. Finally, vacuum oscillations are determined by
the mass differences and vacuum mixing angle $\theta$, so that
\begin{eqnarray}\label{eq:vdef}
 \omega_{\bf p}&=&(m_1^2-m_2^2)/2p\,,\nonumber\\
 {\bf B}&=&(\sin2\theta,0,\cos2\theta)\,.
\end{eqnarray}
Of course, we could have oriented ${\bf B}$ in any other direction in
the $x$-$y$-plane, i.e., it is our choice to set $B_y=0$. For the
normal hierarchy where $m_1<m_2$ the oscillation frequency is
negative. In the main text we prefer to keep a positive $\omega$ which
implies that we have to reverse the $z$-component of ${\bf B}$ for the
normal hierarchy.

The terms proportional to the unit matrix in Eq.~(\ref{eq:matrices})
disappear from the equation of motion Eq.~(\ref{eq:eom1}) due to its
commutator structure, leaving us with the well-known spin-precession
equations
\begin{widetext}
\begin{eqnarray}\label{eq:eom2}
 \partial_t{\bf P}_{\bf p}&=&+\left\{\omega_{\bf p}{\bf B}
 +\sqrt{2}\,G_{\rm F}\left[n_e {\bf L}
 +\int\!\frac{d^3{\bf q}}{(2\pi)^3}
 \left({\bf P}_{\bf q}-\bar{\bf P}_{\bf q}
 \right)(1-\cos\theta_{\bf pq})\right]\right\}
 \times{\bf P}_{\bf p}\,,\nonumber\\
 \partial_t\bar{\bf P}_{\bf p}&=&-\left\{\omega_{\bf p}{\bf B}
 -\sqrt{2}\,G_{\rm F}\left[n_e {\bf L}
 +\int\!\frac{d^3{\bf q}}{(2\pi)^3}
 \left({\bf P}_{\bf q}-\bar{\bf P}_{\bf q}
 \right)(1-\cos\theta_{\bf pq})\right]\right\}
 \times\bar{\bf P}_{\bf p}\,.
\end{eqnarray}
\end{widetext}
In the main text we use the frequency
\begin{equation}
\lambda=\sqrt{2}G_{\rm F}n_e
\end{equation}
as a coefficient for ${\bf L}$ to quantify the matter effect. An
ensemble consisting initially of $\nu_e$ and $\bar\nu_e$ corresponds
to $\int {\bf P}_{\bf p}\,d^3{\bf p}/(2\pi)^3=(0,0,n_{\nu_e})$.
Therefore, if we represent the entire $\nu_e$ ensemble with a single
integrated polarisation vector ${\bf P}$ of unit length, the
$\nu$-$\nu$ term must be of the form $\mu({\bf P}-\bar{\bf
P})\times{\bf P}_{\bf p}$, with
\begin{equation}
\mu=\sqrt{2}G_{\rm F} n_{\nu_e}=\sqrt{2}G_{\rm F} n_{\bar\nu_e}\,.
\end{equation}
We use this frequency to denote the strength of the neutrino self
coupling.

The equations of motion of the entire system, assuming all neutrinos
have the same vacuum oscillation frequency, thus become
\begin{eqnarray}\label{eq:matter10}
 \partial_t{\bf P}&=&\left[+\omega{\bf B}+\lambda{\bf L}
 +\mu\left({\bf P}-\bar{\bf P}\right)\right]\times{\bf P}\,,
 \nonumber\\
 \partial_t\bar{\bf P}&=&\left[-\omega{\bf B}+\lambda{\bf L}
 +\mu\left({\bf P}-\bar{\bf P}\right)\right]\times\bar{\bf P}\,.
\end{eqnarray}
The ${\bf B}$-parts of the Hamiltonian and of the equations of
motion correspond exactly to those of a particle and its
antiparticle with a magnetic moment in the presence of a ${\bf
B}$-field. They have opposite magnetic moments and thus spin-precess
in opposite directions.

The matter term includes an important sign-change in that the particle
and antiparticle have equal energies if their spins are aligned, but
their magnetic moments are anti-aligned. Of course, this sign change
reflects that in the presence of a medium, particles and antiparticles
are affected in opposite manners relative to the vacuum term so that
the usual MSW effect occurs for the normal, but not the inverted mass
hierarchy.

\subsection{Supernova Neutrinos}

Bipolar oscillations are primarily important for neutrinos
streaming off a supernova core. Therefore, we briefly state the
typical parameter values expected in this context. In
numerical simulations of supernova neutrino oscillations, it is
often assumed that $\langle E_{\nu_e}\rangle=11~{\rm MeV}$, $\langle
E_{\bar\nu_e}\rangle=16~{\rm MeV}$ and $\langle
E_{\nu_x}\rangle=25~{\rm MeV}$ for the other
species~\cite{Pastor:2002we, Duan:2006an}. The ``atmospheric''
neutrino mass difference relevant here is $\Delta
m_{23}^2=1.9$--$3.0\times10^{-3}~{\rm eV}^2$. With $\langle
E_{\nu}\rangle=16~{\rm MeV}$, we may thus use
\begin{equation}\label{eq:wsn}
\omega=0.3~{\rm km}^{-1}
\end{equation}
as a typical number. In a supernova one studies the neutrino flavour
evolution as a function of radius from the neutrino sphere, so
it is useful to express all distances in km and all frequencies in
km$^{-1}$.

Moreover, most numerical simulations assume that all neutrino species
are emitted with the same luminosities, with $L_0=10^{51}~{\rm
erg~s}^{-1}$ being a typical choice. The neutrino-sphere radius is
approximately $R_\nu=10$~km. Therefore, a typical neutrino density at
radius $r$ is $n_\nu=L_0/(\langle E_\nu\rangle 4\pi
r^2)=1.04\times10^{32}~{\rm cm}^{-3}\,r_{10}^{-2}$, where
$r_{10}=r/R_\nu=r/10~{\rm km}$. The relevant density for the
calculation of $\mu$ is the difference between the $\nu_e$ and the
$\nu_\mu$ densities.  This amounts to reducing $n_\nu$ by a factor
$16/11-16/25\approx0.81$, a number that reflects the different average
energies of the different species.  Finally, we need to include the
typical angular factor $1-\cos\Theta_{\bf p q}$ between neutrino
trajectories because collinear neutrinos do not cause refractive
effects for each other. This angular effect is approximately taken
into account with the factor
$F=\frac{1}{2}[1-(1-R_\nu^2/r^2)^{1/2}]^2$ used in previous
``one-angle'' numerical studies and originally worked out in
Ref.~\cite{Qian:1994wh}. Altogether we thus find that
\begin{equation}\label{eq:musn}
 \mu=0.3\times10^5~{\rm km}^{-1}
 \left[1-\left(1-r_{10}^{-2}\right)^{1/2}\right] r_{10}^{-2}
\end{equation}
is a reasonable value for simple estimates. Near the neutrino sphere,
$\mu$ is $10^5$ times larger than $\omega$.

The neutrinos streaming off a supernova core are not isotropic so that
the ``multi-angle'' nature of the problem can be important. Still, in
a given radial direction, the problem can be assumed to have axial
symmetry so that different neutrino modes can be classified by their
angle $c_i=\cos\theta_i$, where $\theta_i$ is the angle of the
neutrino momenta relative to the radial direction, i.e., $c_i$
represents all neutrinos streaming in the direction $\theta_i$,
integrated over all azimuthal directions. The coupling strength for
two different ``modes'' $\theta_i$ and $\theta_j$, weighted by the
angular factors, is then proportional to~\cite{Duan:2006an}
\begin{equation}\label{eq:cicj}
1-c_ic_j\,.
\end{equation}
Of course, in an isotropic medium, where $c_i$ and $c_j$ are both
uniformly distributed between $-1$ and $+1$, this term averages
to~1. However, if we consider neutrinos emitted isotropically from a
flat surface, we will have a uniform distribution in the range $0\leq
c_i\leq 1$.  This provides a simple model for a nonisotropic medium.



\onecolumngrid

\vskip2cm

\section{Erratum}

\hbox to\textwidth{\hfil 5 June 2007 \hfil}

\vskip 0.6cm

\twocolumngrid

In Eq.~(46) of ``Self-induced conversion in dense neutrino gases:
Pendulum in flavour space''~\cite{Hannestad:2006nj} we give
\begin{equation}\label{eq:approx}
 \xi\equiv\frac{\omega}{\mu}\bigg|_{\rm transition}
 =\frac{(1-\alpha)^2}{4(1+\alpha)}
\end{equation}
for the transition between the synchronized and bipolar behavior of
the flavor pendulum. Here, $\omega$ is the vacuum oscillation
frequency, $\mu$ quantifies the neutrino-neutrino interaction
strength, and $0\leq\alpha\leq1$ where $\alpha=n_{\bar\nu}/n_{\nu}$.
The neutrino gas is ``dense'' when $\omega/\mu\ll 1$.
Equation~(\ref{eq:approx}) was derived under this assumption and as
such is self-consistent only if $\alpha$ is not too small.

Subsequently it was noted~\cite{Duan:2007mv} that this approximation
is not necessary. In the relevant case of a vanishing mixing angle,
we have $\sigma=1-\alpha$ in Eq.~(45) of
Ref.~\cite{Hannestad:2006nj} and $Q=(1+\alpha-\xi)$ for the inverted
hierarchy so that Eq.~(45) is equivalent to
\begin{equation}
(1-\alpha)^2=4\,\xi\,(1+\alpha-\xi)\,.
\end{equation}
It is solved by
\begin{equation}\label{eq:full}
\xi=\frac{(1-\sqrt\alpha)^2}{2},
\end{equation}
in agreement with Eq.~(76) of Ref.~\cite{Duan:2007mv}.

To compare Eq.~(\ref{eq:approx}) with its alternative form
Eq.~(\ref{eq:full}), we use $\alpha=1-\sigma$ and expand,
\begin{equation}
 \xi=\frac{\sigma^2}{8}\times
 \cases{1+\frac{1}{2}\,\sigma+\frac{1}{4}\,\sigma^2+\ldots
 &from Eq.~(\ref{eq:approx}),\cr
 1+\frac{1}{2}\,\sigma+\frac{5}{16}\,\sigma^2+\ldots&
 from Eq.~(\ref{eq:full}).\cr}
\end{equation}
Both results are surprisingly similar. For the example $\alpha=0.8$
used in Refs.~\cite{Hannestad:2006nj,Duan:2007mv}, the approximate
result from Eq.~(\ref{eq:approx}) is 0.3\% smaller than the exact
one of Eq.~(\ref{eq:full}).

As noted in Ref.~\cite{Duan:2007mv}, Fig.~8 and the text around
Eq.~(49) of our paper are not consistent with the definitions of the
energies in Eqs.~(47) and~(48). In Fig.~8 and its discussion, we
have used $E_{\rm pot}=\omega\,{\bf B}\cdot({\bf P}+\bar{\bf P})$
for the potential energy, i.e., the first part of Eq.~(34), and the
second part of Eq.~(34) for the kinetic energy. In this way the
potential energy does not depend on $\mu$, and the kinetic energy
vanishes for $\mu\to0$. On the other hand, the flavor pendulum is
described by ${\bf Q}={\bf P}+\bar{\bf P}-(\omega/\mu){\bf B}$. Its
potential energy is $\omega{\bf B}\cdot{\bf Q}$. When $\omega/\mu$
is not small, there is an important difference between these
descriptions. For a system with fixed $\mu$ one can add arbitrary
functions of $\mu$ to the potential or to the kinetic energy without
affecting the dynamics. On the other hand, when $\mu$ and thus the
Hamiltonian depend on time, one has to be more careful. The main
point of Fig.~8, the discovery of the approximate equipartition of
energies as the system evolves, remains unaffected and is
analytically elaborated in Ref.~\cite{Duan:2007mv} in terms of
adiabatic invariants.

As noted in Ref.~\cite{Raffelt:2007yz}, another clarification
concerns the normalization of the energies. Equation~(34) and
subsequently all expressions for $E_{\rm pot}$ and $E_{\rm kin}$
should include a factor~1/2 if the energy is to be interpreted as an
energy per neutrino. The quantum Hamiltonian Eq.~(67) is normalized
without ambiguity. The normalization of the classical energy is
inconsequential for our results.



\begin{thebibliography}{00}

\bibitem{Wolfenstein:1977ue}
  L.~Wolfenstein,
  ``Neutrino oscillations in matter,''
  Phys.\ Rev.\ D {\bf 17}, 2369 (1978).

\bibitem{Mikheev:1986gs}
  S.~P.~Mikheev and A.~Y.~Smirnov,
  ``Resonance enhancement of oscillations in matter and solar neutrino
  spectroscopy,''
  Yad.\ Fiz.\  {\bf 42}, 1441 (1985)
  [Sov.\ J.\ Nucl.\ Phys.\  {\bf 42}, 913 (1985)].

\bibitem{Mikheev:1986wj}
  S.~P.~Mikheev and A.~Y.~Smirnov,
  ``Resonant amplification of neutrino oscillations in matter and
  solar neutrino spectroscopy,''
  Nuovo Cim.\ C {\bf 9}, 17 (1986).

\bibitem{Fuller1987}
  G.~M.~Fuller, R.~W.~Mayle, J.~R.~Wilson and D.~N.\ Schramm,
  ``Resonant neutrino oscillations and stellar collapse''
  Astrophys.\ J.\ {\bf 322} (1987) 795.

\bibitem{Notzold:1987ik}
  D.~N\"otzold and G.~Raffelt,
  ``Neutrino dispersion at finite temperature and density,''
  Nucl.\ Phys.\ B {\bf 307}, 924 (1988).

\bibitem{Kuo:1989qe}
  T.~K.~Kuo and J.~T.~Pantaleone,
  ``Neutrino oscillations in matter,''
  Rev.\ Mod.\ Phys.\  {\bf 61}, 937 (1989).

\bibitem{Pantaleone:1992eq}
  J.~Pantaleone,
  ``Neutrino oscillations at high densities,''
  Phys.\ Lett.\ B {\bf 287}, 128 (1992).


\bibitem{Samuel:1993uw}
  S.~Samuel,
  ``Neutrino oscillations in dense neutrino gas\-es,''
  Phys.\ Rev.\ D {\bf 48}, 1462 (1993).

\bibitem{Kostelecky:1993yt}
  V.~A.~Kosteleck\'y, J.~Pantaleone and S.~Samuel,
  ``Neutrino oscillation in the early universe,''
  Phys.\ Lett.\ B {\bf 315}, 46 (1993).

\bibitem{Kostelecky:1993dm}
  V.~A.~Kosteleck\'y and S.~Samuel,
  ``Neutrino oscillations in the early universe with an inverted
  neutrino mass hierarchy,''
  Phys.\ Lett.\ B {\bf 318}, 127 (1993).

\bibitem{Kostelecky:1994ys}
  V.~A.~Kosteleck\'y and S.~Samuel,
  ``Nonlinear neutrino oscillations in the expanding universe,''
   Phys.\ Rev.\ D {\bf 49}, 1740 (1994).

\bibitem{Kostelecky:1995dt}
  V.~A.~Kosteleck\'y and S.~Samuel,
  ``Self-maintained coherent oscillations in dense neutrino gases,''
  Phys.\ Rev.\ D {\bf 52}, 621 (1995)
  [hep-ph/9506262].

\bibitem{Kostelecky:1995xc}
  V.~A.~Kosteleck\'y and S.~Samuel,
  ``Neutrino oscillations in the early universe with nonequilibrium
  neutrino distributions,''
  Phys.\ Rev.\ D {\bf 52}, 3184 (1995)
  [hep-ph/9507427].

\bibitem{Samuel:1996ri}
  S.~Samuel,
  ``Bimodal coherence in dense selfinteracting neutrino gases,''
  Phys.\ Rev.\ D {\bf 53}, 5382 (1996)
  [hep-ph/9604341].

\bibitem{Kostelecky:1996bs}
  V.~A.~Kosteleck\'y and S.~Samuel,
  ``Nonequilibrium neutrino oscillations in the early universe with
  an inverted neutrino mass hierarchy,''
  Phys.\ Lett.\ B {\bf 385}, 159 (1996)
  [hep-ph/9610399].

\bibitem{Pantaleone:1998xi}
  J.~Pantaleone,
  ``Stability of incoherence in an isotropic gas of oscillating
  neutrinos,''
  Phys.\ Rev.\ D {\bf 58}, 073002 (1998).


\bibitem{Pastor:2001iu}
  S.~Pastor, G.~G.~Raffelt and D.~V.~Semikoz,
  ``Physics of synchronized neutrino oscillations caused by
  self-interactions,''
  Phys.\ Rev.\ D {\bf 65}, 053011 (2002)
  [hep-ph/0109035].

\bibitem{Lunardini:2000fy}
  C.~Lunardini and A.~Y.~Smirnov,
  ``High-energy neutrino conversion and the lepton asymmetry in the
  universe,''
  Phys.\ Rev.\ D {\bf 64}, 073006 (2001)
  [hep-ph/0012056].

\bibitem{Dolgov:2002ab}
  A.~D.~Dolgov, S.~H.~Hansen, S.~Pastor, S.~T.~Petcov,
  G.~G.~Raffelt and D.~V.~Semikoz,
  ``Cosmological bounds on neutrino degeneracy improved by flavor
  oscillations,''
  Nucl.\ Phys.\ B {\bf 632}, 363 (2002)
  [hep-ph/0201287].

\bibitem{Wong:2002fa}
  Y.~Y.~Y.~Wong,
  ``Analytical treatment of neutrino asymmetry equilibration
  from flavour oscillations in the early universe,''
  Phys.\ Rev.\ D {\bf 66}, 025015 (2002)
  [hep-ph/ 0203180].

\bibitem{Abazajian:2002qx}
  K.~N.~Abazajian, J.~F.~Beacom and N.~F.~Bell,
  ``Stringent constraints on cosmological neutrino antineutrino
  asymmetries from synchronized flavor transformation,''
  Phys.\ Rev.\ D {\bf 66}, 013008 (2002)
  [astro-ph/0203442].


\bibitem{Duan:2005cp}
  H.~Duan, G.~M.~Fuller and Y.~Z.~Qian,
  ``Collective neutrino flavor transformation in supernovae,''
  astro-ph/0511275.

\bibitem{Duan:2006an}
  H.~Duan, G.~M.~Fuller, J.~Carlson and Y.~Z.~Qian,
  ``Simulation of coherent non-linear neutrino flavor transformation
  in the supernova environment. I: Correlated neutrino trajectories,''
  astro-ph/0606616.

\bibitem{Duan:2006jv}
  H.~Duan, G.~M.~Fuller, J.~Carlson and Y.~Z.~Qian,
  ``Coherent development of neutrino flavor in the supernova
  environment,''
  astro-ph/0608050.

 \bibitem{Keil:2002in}
  M.~T.~Keil, G.~G.~Raffelt and H.~T.~Janka,
  ``Monte Carlo study of supernova neutrino spectra formation,''
  Astrophys.\ J.\  {\bf 590}, 971 (2003)
  [astro-ph/0208035].


\bibitem{Pantaleone:1994ns}
  J.~T.~Pantaleone,
  ``Neutrino flavor evolution near a supernova's core,''
  Phys.\ Lett.\ B {\bf 342}, 250 (1995)
  [astro-ph/9405008].

\bibitem{Qian:1994wh}
  Y.~Z.~Qian and G.~M.~Fuller,
  ``Neutrino-neutrino scattering and matter enhanced neutrino flavor
  transformation in Supernovae,''
  Phys.\ Rev.\ D {\bf 51}, 1479 (1995)
  [astro-ph/9406073].

\bibitem{Sigl:1994hc}
  G.~Sigl,
  ``Neutrino mixing constraints and supernova nucleosynthesis,''
  Phys.\ Rev.\ D {\bf 51}, 4035 (1995)
  [astro-ph/9410094].

\bibitem{Pastor:2002we}
  S.~Pastor and G.~Raffelt,
  ``Flavor oscillations in the supernova hot bubble region:
  Nonlinear  effects of neutrino background,''
  Phys.\ Rev.\ Lett.\  {\bf 89}, 191101 (2002)
  [astro-ph/0207281].

\bibitem{Balantekin:2004ug}
  A.~B.~Balantekin and H.~Y\"uksel,
  ``Neutrino mixing and nucleosynthesis in core-collapse supernovae,''
  New J.\ Phys.\  {\bf 7}, 51 (2005)
  [astro-ph/0411159].


\bibitem{Friedland:2003eh}
  A.~Friedland and C.~Lunardini,
  ``Do many-particle neutrino interactions cause a novel
  coherent effect?,''
  JHEP {\bf 0310}, 043 (2003)
  [hep-ph/0307140].

\bibitem{Friedland:2006ke}
  A.~Friedland, B.~H.~J.~McKellar and I.~Okuniewicz,
  ``Construction and analysis of a simplified many-body
  neutrino model,''
  Phys.\ Rev.\ D {\bf 73}, 093002 (2006)
  [hep-ph/0602016].

\bibitem{Balantekin:2006tg}
  A.~B.~Balantekin and Y.~Pehlivan,
  ``Neutrino neutrino interactions and flavor mixing
  in dense matter,''
  astro-ph/0607527.

\bibitem{Friedland:2003dv}
  A.~Friedland and C.~Lunardini,
  ``Neutrino flavor conversion in a neutrino background:
  Single- versus multi-particle description,''
  Phys.\ Rev.\ D {\bf 68}, 013007 (2003)
  [hep-ph/0304055].

\bibitem{Bell:2003mg}
  N.~F.~Bell, A.~A.~Rawlinson and R.~F.~Sawyer,
  ``Speed-up through entanglement: Many-body effects
  in neutrino processes,''
  Phys.\ Lett.\ B {\bf 573}, 86 (2003)
  [hep-ph/0304082].

\bibitem{Qian:1993dg}
  Y.~Z.~Qian, G.~M.~Fuller, G.~J.~Mathews, R.~Mayle, J.~R.~Wilson
  and S.~E.~Woosley,
  ``A connection between flavor mixing of cosmologically significant
  neutrinos and heavy element nucleosynthesis in supernovae,''
  Phys.\ Rev.\ Lett.\  {\bf 71}, 1965 (1993).

\bibitem{Fuller1992}
  G.~M.~Fuller, R.~Mayle, B.~S.~Meyer and J.~R.~Wilson,
  ``Can a closure mass neutrino help solve the supernova shock
  reheating problem?'',
  Astrophys.\ J.\ {\bf 389}, 517 (1992).



\bibitem{Schirato:2002tg}
  R.~C.~Schirato and G.~M.~Fuller,
  ``Connection between supernova shocks, flavor transformation,
  and the neutrino signal,''
  astro-ph/0205390.

\bibitem{Takahashi:2002yj}
  K.~Takahashi, K.~Sato, H.~E.~Dalhed and J.~R.~Wilson,
  ``Shock propagation and neutrino oscillation in supernova,''
  Astropart.\ Phys.\  {\bf 20}, 189 (2003)
  [astro-ph/0212195].

\bibitem{Lunardini:2003eh}
  C.~Lunardini and A.~Y.~Smirnov,
  ``Probing the neutrino mass hierarchy and the 13-mixing with
  supernovae,''
  JCAP {\bf 0306}, 009 (2003)
  [hep-ph/0302033].

\bibitem{Fogli:2003dw}
  G.~L.~Fogli, E.~Lisi, D.~Montanino and A.~Mirizzi,
  ``Analysis of energy- and time-dependence of supernova shock
  effects on neutrino crossing probabilities,''
  Phys.\ Rev.\ D {\bf 68}, 033005 (2003)
  [hep-ph/0304056].

\bibitem{Tomas:2004gr}
  R.~Tom\`as, M.~Kachelrie{\ss}, G.~Raffelt, A.~Dighe, H.~T.\ Janka
  and L.~Scheck,
  ``Neutrino signatures of supernova shock and reverse shock
  propagation,''
  JCAP {\bf 0409}, 015 (2004)
  [astro-ph/0407132].

\bibitem{Dasgupta:2005wn}
  B.~Dasgupta and A.~Dighe,
  ``Phase effects in neutrino conversions during a supernova
  shock wave,''
  hep-ph/0510219.

\bibitem{Fogli:2004ff}
  G.~L.~Fogli, E.~Lisi, A.~Mirizzi and D.~Montanino,
  ``Probing supernova shock waves and neutrino flavor transitions in
  next-generation water-Cherenkov detectors,''
  JCAP {\bf 0504}, 002 (2005)
  [hep-ph/0412046].

\bibitem{Choubey:2006aq}
  S.~Choubey, N.~P.~Harries and G.~G.~Ross,
  ``Probing neutrino oscillations from supernovae shock waves via
  the IceCube detector,''
  hep-ph/0605255.

\bibitem{Fogli:2006xy}
  G.~L.~Fogli, E.~Lisi, A.~Mirizzi and D.~Montanino,
  ``Damping of supernova neutrino transitions in stochastic shock-wave
  density profiles,''
  JCAP {\bf 0606}, 012 (2006)
  [hep-ph/0603033].


\bibitem{Dighe:1999bi}
  A.~S.~Dighe and A.~Y.~Smirnov,
  ``Identifying the neutrino mass spectrum from the neutrino burst
  from a supernova,''
  Phys.\ Rev.\ D {\bf 62}, 033007 (2000)
  [hep-ph/9907423].

\bibitem{Lunardini:2001pb}
  C.~Lunardini and A.~Y.~Smirnov,
  ``Supernova neutrinos: Earth matter effects and neutrino mass
  spectrum,''
  Nucl.\ Phys.\ B {\bf 616}, 307 (2001)
  [hep-ph/0106149].

\bibitem{Dighe:2003be}
  A.~S.~Dighe, M.~T.~Keil and G.~G.~Raffelt,
  ``Detecting the neutrino mass hierarchy with a supernova at
  IceCube,''
  JCAP {\bf 0306}, 005 (2003)
  [hep-ph/0303210].

\bibitem{Chiu:2006}
  S.~H.~Chiu and T.~K.~Kuo,
  ``Probing neutrino mass hierarchies and $\Theta_{13}$ with
  supernova neutrinos,''
  Phys.\ Rev.\ D {\bf 73}, 033007 (2006);
  Erratum ibid.\ D {\bf 73}, 059901 (2006).

\bibitem{Mirizzi:2006xx}
  A.~Mirizzi, G.~G.~Raffelt and P.~D.~Serpico,
  ``Earth matter effects in supernova neutrinos:
  Optimal detector locations,''
  JCAP {\bf 0605}, 012 (2006)
  [astro-ph/0604300].


\bibitem{Kachelriess:2004ds}
  M.~Kachelrie{\ss}, R.~Tom\`as, R.~Buras, H.~T.~Janka, A.~Marek
  and M.~Rampp,
  ``Exploiting the neutronization burst of a galactic supernova,''
  Phys.\ Rev.\ D {\bf 71}, 063003 (2005)
  [astro-ph/0412082].


\bibitem{Dolgov:1980cq}
  A.~D.~Dolgov,
  ``Neutrinos in the early universe,''
  Yad.\ Fiz.\  {\bf 33}, 1309 (1981)
  [Sov.\ J.\ Nucl.\ Phys.\  {\bf 33}, 700 (1981)].

\bibitem{Sigl:1992fn}
  G.~Sigl and G.~Raffelt,
  ``General kinetic description of relativistic mixed neutrinos,''
  Nucl.\ Phys.\ B {\bf 406}, 423 (1993).

\bibitem{mckellar&thomson}
B.~H.~J.~McKellar and M.~J.~Thomson,
  ``Oscillating doublet neutrinos in the early universe,''
  Phys.\ Rev.\ D {\bf 49}, 2710 (1994).

\end{thebibliography}

\begin{thebibliography}{00}


\bibitem{Hannestad:2006nj}
  S.~Hannestad, G.~G.~Raffelt, G.~Sigl and Y.~Y.~Y.~Wong,
  ``Self-induced conversion in dense neutrino gases: Pendulum in flavour
  space,''
  Phys.\ Rev.\  D {\bf 74}, 105010 (2006)
  [arXiv:astro-ph/0608695].

\bibitem{Duan:2007mv}
  H.~Duan, G.~M.~Fuller, J.~Carlson and Y.~Z.~Qian,
  ``Analysis of collective neutrino flavor transformation in supernovae,''
  arXiv:astro-ph/0703776.

\bibitem{Raffelt:2007yz}
  G.~G.~Raffelt and G.~Sigl,
  ``Self-induced decoherence in dense neutrino gases,''
  Phys.\ Rev.\  D {\bf 75}, 083002 (2007)
  [arXiv:hep-ph/0701182].

\end{thebibliography}
\end{document}